\documentclass[twocolumn]{aastex631}

\usepackage{amsmath}
\usepackage{multirow}

\newcommand{\imig}{J.~Imig ({\it in prep})\,}

\begin{document}

\title{Watching our Galaxy Grow Up: \\ The Mass and Color Evolution of the Milky Way}

\author[0000-0001-6761-9359]{Gail Zasowski}
\affiliation{Department of Physics \& Astronomy, University of Utah, Salt Lake City, UT 84112, USA}

\author[0000-0003-2025-3585]{Julie Imig}
\affiliation{Space Telescope Science Institute, Baltimore, MD 21218, USA}

\author[0009-0007-9715-9147]{Hayley Coluccio}
\affiliation{Department of Physics \& Astronomy, University of Utah, Salt Lake City, UT 84112, USA}



\begin{abstract}

Using our rich observations within the Milky Way to better understand galaxy evolution requires understanding what the Milky Way looks like ``as a galaxy'' -- that is, its ``true'' shape and abundance profiles (unskewed by observational biases), signatures of past mergers and significant accretion events, and even its total stellar mass and integrated SED, which have historically been difficult to constrain. We present a new approach to determining the Milky Way's integrated mass and colors, using recent measurements of the intrinsic density profiles of stellar populations spanning nearly 13~Gyr in time and 1.5~dex in metallicity (representing nearly all of the Galaxy's stars). We trace the evolution of the Milky Way in various diagnostic spaces, explore the impact of specific events on the present-day Milky Way's integrated properties, and use TNG50 simulations to identify ``young'' Galactic analogs and their eventual fates, compared to the real Milky Way's path. 
From the simulation comparisons, we find strong evidence for an earlier-than-average stellar mass assembly of the MW, and that present-day MW analogs follow a similar growth history, albeit at slightly later times; we also find that analogs of the early MW are in no way guaranteed to follow the MW's subsequent path.
This empirical study offers new constraints on our ``Galaxy as a galaxy'' --- today and across cosmic time -- and on its place in the general galactic population. 

\end{abstract}



\section{Introduction} 
\label{sec:intro}

For nearly all of human history, our perspective of our home galaxy (long before such a concept was defined) was that it spanned the entire universe. Even as our understanding of the concept of galaxies evolved --- from the ``spiral nebulae'' revealed by increasingly sophisticated  telescopes \citep[e.g.,][]{Rosse_1850_spiralnebulae}, to the discovery by \citet{Hubble_1929_m31} and others that the Milky Way is simply one of those many nebulae --- 
our knowledge of what our home galaxy would look like as one of those distant systems has not kept pace.

The Milky Way presents a unique opportunity to study galaxy evolutionary processes in incredibly spectacular detail --- but specifically, in a single $L \approx L^*$ spiral galaxy at $z=0$ \citep[e.g.,][]{BlandHawthorn_2016_MWreview}. 
For more than a century, detailed studies of the Sun and nearby and/or bright stars have shaped our understanding of the inner workings of stars and stellar evolution \citep[e.g.,][]{Russell_1914_hrd,Schou_1998_solarinterior,Filippazzo_2015_starSEDs,Kurtz_2022_asteroseismology}. Recent and ongoing surveys of larger numbers of stars have allowed us to test that understanding in a wide range of galactic environments, improve where needed, and apply that knowledge to understanding the structure, dynamics, chemistry, and evolution of the Milky Way's ecosystem\footnote{This is a paper about stars, but certainly much the same is true for the myriad of atomic gas, molecular gas, and dust studies that are (literally) filling in the gaps between the stars.}. These surveys include both photometric \citep[e.g., 2MASS, DECAPS, PanSTARRS;][]{Skrutskie_06_2mass,Schlafly_2018_decaps1,Saydjari_2023_decaps2,Chambers_2016_panstarrs1} and spectroscopic \citep[e.g., LAMOST, APOGEE, GALAH;][]{Cui_2012_lamost,Majewski_2017_apogeeoverview,DeSilva_2015_galah} efforts, which in turn enable remarkable advances in, e.g., astrometry \citep[e.g.,][]{GaiaCollab_2023_dr3} and asteroseismology/age-dating \citep[e.g.,][]{Anders_2017_corotgee,Pinsonneault_2018_apokasc2}. 

At the same time, extragalactic studies have shown us how galaxy evolution can play out in both strikingly similar and very different ways in systems of different masses, environments, star formation rates, and so on. Targeted studies of smaller numbers of relatively nearby systems provide highly-resolved information at recent times (though that horizon is constantly increasing!) \citep[e.g., CALIFA and MaNGA;][]{Sanchez_2012_califa,Bundy_2015_manga}. Complementary large galaxy surveys provide a time-stamped progression of, for example, the mean and variance in integrated galaxy properties back across cosmic time \citep[e.g., SDSS and DESI;][]{York_2000_sdss,Dey_2019_DESIsurveys}.

Galaxy evolution simulations are similarly tiered \citep{Vogelsberger_2020_galaxysims}, spanning dark-matter-only systems on cosmological scales \citep[e.g., Millennium-XXL;][]{Angulo_2012_millenniumxxl}, to individual galaxy simulations that incorporate detailed baryonic physics at sub-kpc scales \citep[e.g., Latte/FIRE or Auriga;][]{Wetzel_2016_Latte,Grand_2017_auriga}. No single project can include the full range of both size and physics, but very exciting progress has been made toward ``zooming in'' on cosmologically-scaled results in a way unavailable to observers \cite[e.g.,][]{Barnes_2017_clustereagle}.

One key piece missing, however, is a precise measurement of the integrated characteristics of the MW itself --- that is, how it would be characterized in those large galaxy surveys, relative to other systems. The primary reason for this is the same reason we understand local stars and stellar populations so well: the Sun is deeply embedded in the disk of the MW, where obscuring dust \citep{Bovy_2016_MWdust3D} and distance uncertainties confound our ability to infer a global picture and evolutionary history. Even such basic properties as the MW's total stellar mass and integrated colors, which feature in a large number of fundamental galaxy relations, have large uncertainties relative to those of other galaxies --- both random uncertainties due to the difficulty of measuring these quantities internally, and systematic uncertainties due to the very different methods required to measure them. And because we have only a snapshot of our galaxy at the present day, inferring how these fundamental parameters have changed over time is even more challenging.

This is a problem well worth solving, not only to avoid over-tuning simulations to fit the well-measured small-scale features \citep[e.g.,][]{Fielder_2021_mwSED}, but also because the MW has long been suspected to be a slight outlier in those global properties we {\it have} been able to estimate.

Previous work paints a picture of a multi-armed, semi-flocculent barred spiral galaxy with an exponential scale length that places it among the most compact galaxies of its type and stellar mass \citep[e.g.,][but see also \citealp{Frankel_2019_MWgrowth} and \citealp{Lian_2024_MWsize2} for contrasting measurements]{Reid_2014_masers,Portail_2015_M2Mbulgemodels,Licquia_2016_MWscalingrelations,Boardman_2020_MWAsM31As}. It has a relatively low star formation rate (SFR) of $\sim$2--3~M$_\odot$~yr$^{-1}$ \citep[e.g.,][]{Licquia_2015_MWmassSFR,FraserMcKelvie_2019_MWsfr,Elia_2022_MWsfr,Zari_2023_MWsfr}, which is likely at least part of the reason that estimates of its integrated colors place it in the so-called ``green valley'' in the galaxy mass--color plane \citep[e.g.,][]{Mutch_2011_greenvalley,Licquia_2015_MWcolorL,Fielder_2021_mwSED,Natale_2022_allMW}. The green valley is a relatively underpopulated region in this plane that lies between the bluer, generally lower-mass, star-forming main sequence, and the cloud of redder, generally higher-mass, passive galaxies \citep[e.g.,][]{Baldry_2006_bimodalgalaxies}. As a normal (if sometimes short-lived) stage in the evolution of a massive galaxy, this characterization suggests that the MW, while not a worryingly distinct outlier from our general picture of galaxy evolution, is not an archetypal example of a generic spiral galaxy.

And what about how the MW got to this state? Has it always been a little bit ``off'' the fundamental relations defined by its galactic contemporaries, or did it undergo an unusual stochastic event or two that pushed it away? The history of our Galaxy is necessarily inferred indirectly, using, for example, the current properties of stars with age estimates (of varying reliability) or simulations of presumed analogs at higher redshift. For instance, \citet{Mackereth_2018_simalphaelements} used $\alpha$-element abundance distributions in EAGLE simulations, in comparison to that of the MW, to argue that the MW experienced a more rapid {\it early} accretion of gas/stellar mass than many seemingly similar galaxies. This basic picture has been enhanced by studies of stellar kinematical--chemical patterns that suggest the MW's disk --- at least, the inner, thicker part --- formed extremely early \citep[often using metallicities as proxies for age; e.g.,][]{Belokurov_2022_aurora,Rix_2022_pooroldheart,Chandra_2024_3phaseMW,Thulasidharan_2024_agethickness,Chen_2025_z2starburst}. Both the event and its rarity have been supported by some simulations \citep[e.g.,][]{Semenov_2024_MWdisk}. After this early growth, however, the MW appears to have undergone a gentler and less eventful interaction history than similar systems \citep[e.g.,][]{Hammer_2007_MWtooquiet,Bonaca_2020_earlyMW}.

However, through all of this, we must keep in mind that estimates of the MW's integrated properties --- including its total luminosity and colors, stellar mass, disk size, star formation history, merger history, and light-weighted abundances --- are largely based either i) on very different calculations of these quantities than in the galaxies to which we compare the MW, or ii) on other galaxies that are deemed sufficiently similar to the MW to be a representative predictive sample. One common question between these is simply: have we {\it measured} enough of the MW to be confident we can describe all of it? Until the arrival of large surveys capable of probing both deeply and broadly across the disk and bulge, the answer was generally ``no''. But now that is shifting: 
the ability to make not only ``all-MW'' measurements, but also estimates of what those measurements would have been in the past, is increasingly feasible.

In this paper, we demonstrate a few examples of the power of such a time-resolved view of the integrated MW. In \S\ref{sec:data}, we describe our approach to computing the stellar mass and integrated magnitudes/colors of the MW, and in \S\ref{sec:mw_today} we explore what could be done with such measurements in terms of metallicities and star formation histories. In \S\ref{sec:time_machine}, we describe the method of ``de-aging'' the MW to earlier times and making similar integrated measurements, and in \S\ref{sec:mergers}, we check the impact of including two of the largest past merger events (the Gaia-Enceladus progenitor and the Sagittarius dwarf galaxy). Finally, in \S\ref{sec:sim_analogs}, we turn to simulations and explore not only systems that resemble the MW today, but also the evolution of analogs of the younger MW, to demonstrate how this approach could be used to better understand the evolution of the MW itself across cosmic time. We summarize in \S\ref{sec:conclusions}.

\section{Recipe for an Integrated Milky Way}
\label{sec:data}

\subsection{Stellar Data from APOGEE}
\label{sec:apogee}

The Apache Point Observatory Galactic Evolution Experiment \citep[APOGEE;][]{Majewski_2017_apogeeoverview}, a stellar spectroscopic survey that spanned both SDSS-III and SDSS-IV \citep{Eisenstein_11_sdss3overview,Blanton_2017_sdss4}, collected high-resolution ($R \sim 22,500$), near-infrared ($\lambda = 1.5-1.7$~$\mu$m) spectra for $\sim$657,000 stars throughout the Milky Way and Local Group galaxies \citep{Zasowski_2013_apogeetargeting,Zasowski_2017_apogee2targeting,Beaton_2021_apogee2Ntargeting,Santana_2021_apogee2Stargeting}. Data were taken with near-twin 300-fiber spectrographs \citep{Wilson_2019_apogeespectrographs} linked to the 2.5~m Sloan Foundation Telescope \citep{Gunn_2006_sloantelescope} at the Apache Point Observatory and the 2.5~m Ir\'{e}n\'{e}e du~Pont Telescope at Las Campanas Observatory \citep{Bowen_1973_duPontTelescope}. 

APOGEE's data reduction processes are described in \citet{Nidever_2015_apogeereduction}, and the standard analysis pipelines, linelists, and other components to produce stellar parameters and chemical abundances are described in \citet{GarciaPerez_2016_aspcap,Jonsson_2020_apogeeDR16,Smith_2021_apogeeDR16linelist,Osorio_2020_apogeeNLTE}. A number of VACs based on APOGEE data are also available\footnote{\url{https://www.sdss4.org/dr17/data_access/value-added-catalogs/}}. The final data release of APOGEE data was part of SDSS-IV Data Release~17 in December~2021 \citep{Abdurrouf_2021_sdssDR17}. 

\subsection{Mono-Age and -Abundance Populations}
\label{sec:maaps}

The initial foundation of this work is the suite of stellar density profiles of Milky Way ``mono-age and -abundance populations'' (MAAPs) by \imig, and explained thoroughly in that paper. In summary, more than 200,000 red giant stars with spectra and derived parameters from SDSS-IV/APOGEE DR17 \citep[\S\ref{sec:apogee};][]{Majewski_2017_apogeeoverview,Abdurrouf_2021_sdssDR17}, and distances and ages from the \texttt{DistMass} catalog\footnote{\url{https://www.sdss4.org/dr17/data_access/value-added-catalogs/?vac_id=distmass:-distances,-masses,-and-ages-for-apogee-dr17}} \citep{StoneMartinez_2024_distmass}, were used to compute stellar density profiles as a function of Galactic radius $R_{\rm GC}$, distance from the midplane $Z_{\rm GC}$, stellar age $\tau$, stellar metallicity [M/H], and stellar alpha-element abundance [$\alpha$/M]. 

The MAAPs are defined on a grid spanning 
\begin{itemize} \itemsep -2pt
    \item metallicity: $\rm -1 \le [M/H] \le +0.5$, with spacing $\rm \Delta [M/H] = 0.1$~dex,
    \item age: $9.0 \le \log{\left(\frac{\tau}{\rm yr}\right)} \le 10.1$, with spacing $\rm \Delta \log{\left(\frac{\tau}{\rm yr}\right)} = 0.1$~dex, and 
\end{itemize}
duplicated into ``high'' and ``low'' alpha populations following the division in \citet{Patil_2023_copulas}, for a total of 360 possible MAAPs. Of these, $\sim$170 bins within these grid boundaries ($\sim$48\%) did not have sufficient stars for a reliable density profile fit. The \texttt{DistMass} ages themselves were only used for stars with $\rm [M/H] \ge -0.65$; for high-$\alpha$ MAAPs with lower metallicity, the age distribution was assumed to be the same as at $\rm -0.65 < [M/H] < -0.3$, which themselves share the same age distribution within the uncertainties. (There are so few low-$\alpha$ stars with $\rm [M/H] < -0.65$ that those MAAPs do not carry significant mass anyway.) 

According to the Milky Way synthesis model \texttt{Galaxia} \citep{Sharma_2011_galaxia}, \textcolor{black}{these age and metallicity limits encompass $\gtrsim$95\% of the stars in the Milky Way.} We address the presence of young stars, with $\tau < 1$~Gyr, in \S\ref{sec:young_maaps}.
We also compute an [$\alpha$/M] value for each MAAP, as the median [$\alpha$/M] of the stars in that MAAP --- i.e., within the [M/H] and age bounds, and above/below the \citet{Patil_2023_copulas} division for high/low alpha MAAPs, respectively.

The APOGEE DR17 selection function, as a function of stellar Galactic position, color, and metallicity, is modeled as an inhomogeneous Poisson point process \citep[e.g.,][]{Bovy_2012_diskpops,Bovy_2016_MWdust3D,Mackereth_2017_diskagemetallicity}. 
See \imig for a detailed description.

Each density profile is parameterized as the product of a broken exponential in the radial direction, with the break radius as a free parameter in the fit, and a single exponential in the vertical direction, with a flare parameter that varies linearly with radius \citep[see also][]{Bovy_2016_diskstructure,Mackereth_2017_diskagemetallicity,Yu_2021_lamostRCs,Lian_2022_MWtomography}. Altogether, each profile is fully described by six parameters: a break radius ($R_{\rm break}$), a radial scalelength inside and outside of $R_{\rm break}$ ($h_{R,in}$ and  $h_{R,out}$, respectively), a vertical scaleheight at the solar circle ($h_{Z_\odot}$), a flaring parameter $A_{\rm flare}$, and a scaling parameter to set the measured number density at the solar circle ($\nu_{R_\odot}$). 
These parameters are expressed in components that depend on Galactic radius ($\Sigma(R_{\rm GC})$) and both radius and height above the midplane ($\xi(R_{\rm GC},Z_{\rm GC})$), as follows: 

\begin{align}
\label{eqn:density_profile}
    \nu_*(R_{\rm GC},Z_{\rm GC}) &= \nu_{R_\odot} \Sigma(R_{\rm GC}) \xi(R_{\rm GC},Z_{\rm GC}) \\
    \ln{\Sigma(R)} &\propto 
    \begin{cases}
    -h_{R,\textrm{in}}^{-1}(R-R_{\odot}) - C & ; R \leq R_{\text{break}}\\ \nonumber
    -h_{R,\textrm{out}}^{-1}(R-R_{\odot}) & ; R > R_{\text{break}}\\
    \end{cases} \\ \nonumber
    \ln{\xi(R,Z)} &\propto -h_{Z}(R)^{-1} |Z| \\ \nonumber
    &~ h_{Z}(R) = h_{Z\odot} + A_{\text{flare}}(R - R_{\odot}) \nonumber
\end{align}
As described more fully in Section~3 of \imig, $\nu_{R_\odot}$ is derived using the survey selection function and observed star counts in APOGEE following the integration methods of, e.g., \citet{Bovy_2012_nothickdisk} and \citet{Mackereth_2017_diskagemetallicity}, and the other five parameters are inferred from an MCMC algorithm with a forward-modeling approach.

\begin{figure*}[!hb]
    \centering
    \includegraphics[width=\textwidth]{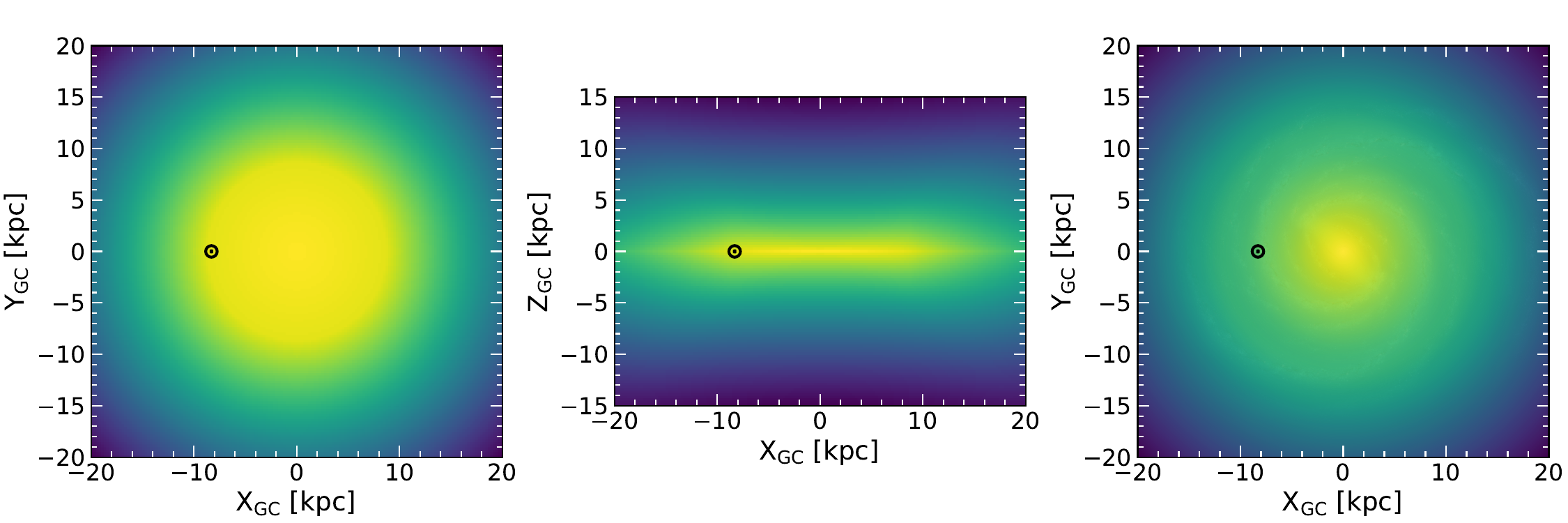}
    \caption{
    {\it Left, center}: A sample MAAP density profile (\S\ref{sec:maaps}), with the parameterization of Eq.\;\ref{eqn:density_profile}. This particular example is for the low-[$\alpha$/M] MAAP with \textcolor{black}{$\rm [M/H] = -0.05$ and $\rm \log{\tau/yr} = 9.75$}, which has $R_{\rm break} = 8.7$~kpc, $h_{\rm R,in} = 5.3$~kpc, $h_{\rm R,out} = 1.1$~kpc, $h_{Z_\odot} = 0.3$~kpc, and $A_{\rm flare} = 0.03$ (see \S\ref{sec:maaps} for the definition of these parameters). The color indicates the log of the number density integrated through $Z_{\rm GC}$ (left) and $Y_{\rm GC}$ (right), and ``$\odot$'' indicates the position of the Sun. {\it Right}: similar to the left-hand projection but for the total density of all MAAPs. Strictly for visual interest, this has been overlaid with Robert Hurt's graphical representation of the Milky Way to highlight the relative placement of the bar and spiral arms.
    }
    \label{fig:example_map_profile}
\end{figure*}

See Figure~\ref{fig:example_map_profile} for an example of such a profile.
These stellar number densities $\nu_*$ (in pc$^{-3}$) represent the density of stars within each MAAP that also satisfy the $3.5 \ge \log{g} \ge 1.0$ criterion imposed by \imig,
and also highlight the broad radial distribution of stars that have presumably migrated from their much narrower range of birth radii \citep[e.g.,][]{Frankel_2018_migration,Lu_2024_birthradii}. To compute the total number of giant stars in this $\log{g}$ range within each MAAP, we evaluate the density profile on a 3D grid of ($X_{\rm GC},Y_{\rm GC},Z_{\rm GC}$) with \textcolor{black}{(0.1~kpc)$^3$} bins, multiply each bin's density by the volume of the bin, and sum the total number of stars. 

Uncertainties are inferred by recalculating this summation in a Monte Carlo fashion. We perturb each of the five MCMC-fitted parameters by a random value drawn from a normal distribution defined by their \textcolor{black}{(two-sided)} uncertainties, repeat the integration 50~times, and explore the resulting median absolute deviation of the counts as the uncertainty in number count for that MAAP.
The median value is $\sim$50\%, with some significantly larger outliers; however, the actual number counts of these outliers are completely consistent with the counts of MAAPs adjacent in age and metallicity, so we hypothesize the seemingly larger count uncertainties are due to overestimated fitting uncertainties. This is further supported by the fact that the fractional standard deviation of MAAP counts within small ranges of $\tau$ and [M/H] is $\sim$30\%, and the fractional median deviation is $\sim$18\%. We thus adopt a constant number count uncertainty of \textcolor{black}{30\% for all MAAPs, and propagate that through to uncertainties on masses and colors in later sections.}

\begin{figure*}[!hp]
    \centering
    \includegraphics[width=\textwidth]{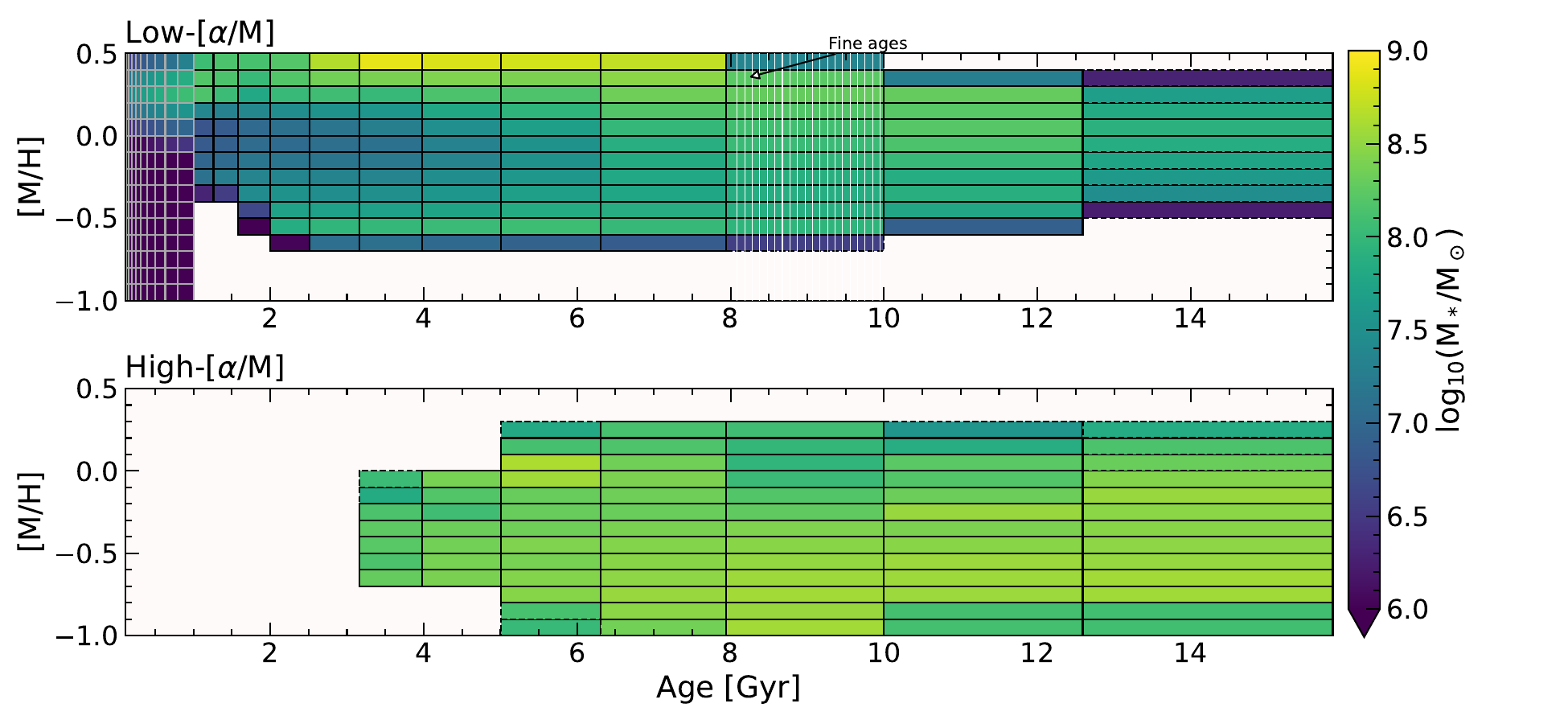}
    \caption{
    MAAP stellar masses in the age--metallicity plane, with low-[$\alpha$/M] populations on top and high-[$\alpha$/M] populations on the bottom. The wide outlined boxes show the span of the original MAAPs, with black solid outlines (original fits from \imig; \S\ref{sec:maaps}), gray solid outlines (the ``young'' MAAPs with $\rm \tau < 1$~Gyr; \S\ref{sec:young_maaps}), or black dotted outlines (interpolated/extrapolated MAAPs; \S\ref{sec:extrapolation}). The thin vertical white lines spanning the low-[$\alpha$/M] 8--10~Gyr boxes indicate the $\sim$0.1~Gyr fine-age spacing (\S\ref{sec:fine_ages}). 
    }
    \label{fig:maap_masses}
\end{figure*}

To convert each MAAP's total number of stars to a MAAP stellar mass, we follow the procedure outlined in \imig. Using PARSEC isochrones\footnote{\label{fn:cmd} \url{http://stev.oapd.inaf.it/cgi-bin/cmd}} \citep[][]{Bressan_2012_parsec,Chen_2014_lowmassPARSEC,Chen_2015_highmassPARSEC,Tang_2014_highmassPARSEC,Marigo_2017_parseccolibri,Pastorelli_2019_tpagb,Pastorelli_2020_tpagb} and assuming a Kroupa IMF \citep[][corrected for unresolved binaries]{Kroupa_2001_imf,Kroupa_2002_imf}, we compute the difference between the \texttt{int\_IMF} value for the minimum- and maximum-mass isochrone points within the relevant $\log{g}$ range of that MAAP's corresponding isochrone. This difference is equal to the number of stars with masses between these two points, per unit mass of the {\it initial} stellar population (given the assumed IMF)\footnote{\url{http://stev.oapd.inaf.it/cmd_3.7/help.html}}. Thus, by dividing the number of inferred giant branch stars in each MAAP by this difference, we calculate the initial MAAP mass represented by that MAAP's APOGEE stars. These masses (slightly refined as in \S\ref{sec:maap_mods} and then corrected for stellar evolution as in \S\ref{sec:different_masses}) are shown in Figure~\ref{fig:maap_masses}, on the grid of age and metallicity.

\subsection{Expanding and Refining the MAAP Coverage}
\label{sec:maap_mods}

\subsubsection{Young MAAPs}
\label{sec:young_maaps}

Due to the selection criteria of APOGEE's ``normal red giant'' sample, which forms the basis of the analysis in \imig and thus this work, very few stars with $\log{\tau} < 9.0$ are represented. However, the MW's stellar nurseries have not been completely idle for the past Gyr, and these populations will have a non-negligible impact on the Galaxy's integrated colors, especially at UV and optical wavelengths.  

Therefore, we extend the age range of the original \imig MAAPs down to 0.1~Gyr with some simple calculations. First, we assume that the star formation rate (SFR) over the past Gyr has been a) constant and b) well-represented by the SFR density profile derived using Herschel/Hi-GAL observations by \citet[][see also \citet{Soler_2023_MWsfr}]{Elia_2022_MWsfr}. We note that the ``dip'' in this $\sum{\rm SFR}$ profile inside $\sim$5~kpc likely has some contribution from incompleteness in the dataset, partially due to inner-Galaxy distance ambiguities. We tested whether artificially enhancing the SFR density at small radii, including up to a flat profile at $R=0-5$~kpc, had any impact on our findings; since it did not, we adopt the \citet{Elia_2022_MWsfr} profile as-is.

Second, we assume that the interstellar metallicity profile also has not changed significantly over the past Gyr, and can be represented during that period by the present-day profile observed in, e.g., 
young Cepheid variables \citep[$-0.06$~dex~kpc$^{-1}$;][]{Genovali_2014_CephMHgrad}, 
$\sim$300-Myr-old MS stars \citep[$-0.02$~dex~kpc$^{-1}$;][]{Wang_2023_metallicitygrad},
HII regions \citep[$-0.04$~dex~kpc$^{-1}$;][]{Balser_2011_HiiMHgrad,Esteban_2017_HiiMHgrad},
OB stars \citep[$-0.04$~dex~kpc$^{-1}$;][]{Daflon_2004_OBMHgrad}, or
Milky Way analog galaxies \citep[$-0.02$~dex~kpc$^{-1}$;][]{Boardman_2020_MWAs}. We opt to use the value of $-0.04$~dex~kpc$^{-1}$ as a rough average of these measurements, but we emphasize that even choosing more extreme values, such as the $-0.07$~dex~kpc$^{-1}$ assumed by \citet{Minchev_2018_MHgrad}, does not affect our findings below.

To calculate the distribution of mass with $\tau < 1$~Gyr as a function of age and metallicity, we 
\begin{enumerate} \itemsep -2pt
    \item construct a series of thin radial rings ($R_{\rm GC}=0-25$~kpc, $\Delta R = 0.1$~kpc),
    \item multiply the interpolated SFR density from \citet{Elia_2022_MWsfr} in each ring by the area of the ring for a total stellar mass at that radius, 
    \item assign to that mass the metallicity inferred for the ``middle'' radial value based on our adopted gradient, 
    \item and divide all of the mass by metallicity among the original MAAP metallicity bins, and within each metallicity, proportionally among log-spaced age bins ($8.0 \le \log{\left(\frac{\tau}{\rm yr}\right)} < 9.0$, $\rm \Delta \log{\left(\frac{\tau}{\rm yr}\right)} = 0.1$~dex).
\end{enumerate}  
These ``young MAAP'' additions to the original set, which contribute a total of \textcolor{black}{$1.7 \times 10^9$~M$_\odot$}, are highlighted in \textcolor{black}{Figure~\ref{fig:maap_masses} by gray outlines}.

\subsubsection{Rescaling the Supersolar Inner Galaxy}

The stellar sample used to fit the MAAP spatial distributions covers a remarkably large range of spatial position ($R_{\rm GC}=0-20$~kpc; \imig). Unsurprisingly, however, due to heavily dust obscuration and low selection fractions, the inner Galaxy is the most underrepresented compared to its actual stellar density. Given this, and the particular analytical form adopted for the density profiles (\S\ref{sec:maaps}), it is likely that the inferred masses for the MAAPs with the highest concentrations in the inner Galaxy are overestimated. 

The solution of adopting a more complex density profile form that includes a bulge/bar is beyond the scope of \imig and this paper. We instead apply a correction by capping the density profile at $R_{\rm GC} = 2.5$~kpc for \textcolor{black}{those MAAPS with $\rm [M/H] = +0.45$ and $\tau > 2$~Gyr}, and using the value at 2.5~kpc for the inner voxels.
\textcolor{black}{As seen in \S\ref{sec:mdfs_and_gradients}, this simple approximation reproduces quite well previous estimates of the inner Galaxy's metallicity distribution.}

\subsubsection{Extrapolation and Smoothing}
\label{sec:extrapolation}

The original \imig density profile parameters are provided only for those MAAPs where sufficient stars were found for a converged fit. This leads to some unphysical ``holes'' and sharp ``edges'' of the mass distribution in the age--metallicity plane, which we attempt to smooth with some simple assumptions. Based on visual inspection of the stellar distribution in the [M/H]--[$\alpha$/M] plane and the number counts in the age--metallicity plane (Figures~2 and 3 in \imig), we set the following ``unfitted'' MAAP masses explicitly to zero\footnote{These limits do not apply to the ``young'' MAAPs generated in \S\ref{sec:young_maaps}.}:
\begin{itemize} \itemsep -2pt
    \item High-[$\alpha$/M]:
    \begin{itemize}
        \item $\rm [M/H] \ge +0.15$, all $\tau$ 
        \item $\rm \log{\tau} < 9.5$, all [M/H] 
        \item $\rm \log{\tau} < 9.7$ and ($\rm [M/H] > -0.15$ or $\rm [M/H] < -0.65$) 
    \end{itemize}
    \item Low-[$\alpha$/M]:
    \begin{itemize}
        \item $\rm [M/H] < -0.65$, all $\tau$ 
        \item $\rm [M/H] < -0.4$, $\log{\tau} < 9.2$
        \item $\rm [M/H] > +0.4$, $\log{\tau} > 10.0$
        \item $\rm [M/H] < -0.5$, $\log{\tau} > 10.1$
    \end{itemize}
\end{itemize}
We emphasize that these MAAPs have always been ``zero'' in the sense of not contributing any mass or luminosity, in \imig or here, but explicitly setting them to zero allows us to estimate non-zero masses for the unfitted-but-probably-not-zero-mass MAAPs in between. We use
\textcolor{black}{\texttt{scipy.ndimage.gaussian\_filter} to smooth the logarithm of the MAAP masses in the 2D age--metallicity plane, with $\sigma=(0.5,0.5)$ for the low-$\alpha$ MAAPs and $\sigma=(0.25,0.25)$ for the high-$\alpha$ MAAPs}. 
These smoothed values are what are shown in Figure~\ref{fig:maap_masses}.

\subsubsection{Fine Age Bins}
\label{sec:fine_ages}

The constant $\Delta \log{\tau}=0.1$ steps in the original MAAP grid from \imig correspond to linear bins of \textcolor{black}{$\sim$2~Gyr at 8~Gyr ago}, and larger at older ages. These steps are driven by the approximately-constant-in-log uncertainties of the ages themselves. However, if these age steps are used as the lookback points explored in \S\ref{sec:time_machine}, the Milky Way's growth becomes increasing ``chunky'' at early times, as large amounts of mass are added at widely-spaced intervals; this in turn results in unphysically sharp changes in color.

Thus, we make the assumption that the Milky Way's historical star formation rate was $\sim$constant {\it within} each $\Delta \log\tau=0.1$ step, and divide the mass in each MAAP evenly between linear age steps of $\Delta\tau \approx 0.1$~Gyr spanning the width of the original age bin. (Due to the uneven size of the original bins in linear age, the actual width of these ``fine'' age bins ranges from 0.09--0.11~Gyr.) This step results in \textcolor{black}{$\sim$3300} ``fine-age'' MAAPs that are used for the rest of this study.

\subsection{From Masses to Magnitudes}
\label{sec:maap_mags}

To calculate magnitudes for our MAAPs, we use the ``integrated magnitudes for single-burst stellar populations'' based on the PARSEC isochrones. 
We downloaded magnitudes for the SDSS {\it ugriz} filters for all combinations of MAAP [M/H] and age included in our sample (and all age steps needed for the lookback process described in \S\ref{sec:time_machine}). These represent absolute magnitude per initial solar mass of a given Simple Stellar Population (SSP; $M_{\lambda,{\rm SSP}}$), so the total absolute magnitude for a MAAP with stellar mass $M_{\rm MAAP}$ is:
\begin{equation} \label{eqn:maap_mags}
    M_{\lambda,0} = M_{\lambda,{\rm SSP}} - 2.5 \log_{10}\left(\frac{M_{\rm MAAP}}{M_\odot}\right).
\end{equation}
Because of the way the MAAP stellar masses are inferred from the stellar number counts (\S\ref{sec:maaps}), these magnitudes represent the brightness of each MAAP at its metallicity and present-day age, including the effects of stellar evolution (see more in \S\ref{sec:different_masses}).

\subsection{Reddening}
\label{sec:reddening}

\textcolor{black}{In \imig, reddening is applied by assuming that half of the mass in each MAAP lies behind a constant $A_V=1$~mag slab of dust, as a simplified approximation of a face-on perspective of the disk.} Here, we make the same assumption that half of each MAAP's mass is unaffected by dust. However, for the other half of the mass, we instead adopt the exponential dust profile used in, e.g., the \texttt{Galaxia} model \citep[][]{Sharma_2011_galaxia,Robin_03_besanconmodel}:
\begin{equation}
    \rho_{\rm dust}(R,\phi,z) = \frac{\rho_0}{k_{\rm flare}} \exp{\left(-\frac{R-R_\odot}{h_{\rm R}}\right)} \times \exp{\left(-\frac{|z-z_{\rm warp}|}{k_{\rm flare}h_{\rm z}}\right)},
\end{equation}
where $\rho_0$ is the volume density of dust at the solar circle ($R=R_\odot$), $h_R$ (4.5~kpc) and $h_z$ (0.14~kpc) are the scalelength and scaleheight of the dust distribution, respectively, and the flare amplitude $k_{\rm flare}$ and vertical warp offset $z_{\rm warp}$ are themselves functions of both $R$ and $\phi$. We limit ourselves to calculating the MW's integrated colors as would be measured face-on, so the vertical structure (and its azimuthal dependence) can be neglected, effectively by setting $k_{\rm flare}=1$~kpc and $z_{\rm warp}=0$~kpc. 

We calculate the fraction of each MAAP's mass in radial bins of $\Delta R_{\rm GC}=0.5$~kpc, and calculate the integrated (face-on) dust extinction in those same bins, scaling the dust density to produce $E(B-V)=0.06$ at the solar circle. This value was chosen as twice the total $E(B-V)$ predicted by \texttt{Galaxia} towards the North or South Galactic Poles \citep{Sharma_2011_galaxia}, since those predictions assume the observation is being made from the middle of disk midplane. 

We assume that the extincted half of each MAAP's mass at a given Galactic radius is behind a single amount of dust, equivalent to the  sightline extinction at that radius integrated through the entire face-on plane. Using the extinction law conversion coefficients $A_\lambda/A_V$ tabulated in \citet[][assuming $R_V=3.1$]{Schlafly_2011_calibSFD}, each MAAP's integrated magnitudes from \S\ref{sec:maap_mags} are extincted by taking into account these coefficients and the fraction of the MAAP's mass ($f_i$) at each radius ($R_i$):
\begin{equation}
    A_\lambda = -2.5 \log{\left(0.5 + 0.5\sum_i f_i 10^{-0.4A_\lambda(R_i)}\right)}.
\end{equation}

We also apply additional extinction to the youngest MAAPs in our sample. The embedding dust of a young cluster is expected to disperse within tens of Myr, along with the stars themselves \citep[e.g.][]{Krumholz_2019_clusters}, but those stars will remain within the dustier-than-average star-forming environment (e.g., a spiral arm) for a longer period. We approximate this time as a few crossing times across a spiral arm, assuming an arm width of a few hundred~pc \citep[e.g.,][]{Reid_2014_masers} and an internal cluster velocity dispersion of a few km~s$^{-1}$ \citep[e.g.,][]{Kuhn_2019_clusterkin}, leading to a value of $\sim$500~Myr. Based on extinction maps in nearby spiral galaxies \citep[e.g.,][]{Dalcanton_2015_PHATextinction,Kahre_2018_EGextinction}, we estimate that a typical star-forming spiral arm's face-on integrated extinction spans $A_V \sim 0.5-1$~mag. Thus, we apply a constant \textcolor{black}{$A_V = 0.75$~mag} of extinction (converted to the {\it ugriz} filters, as above) to the magnitudes of all MAAPs with \textcolor{black}{$\rm \tau \le 0.5$~Gyr}, in addition to the large-scale dust disk extinction.

\subsection{Putting It All Together}
\label{sec:combine_mags}
Finally, to compute the integrated magnitudes of the present-day MW, the absolute magnitudes of all MAAPs are combined:
\begin{equation}
\label{eqn:total_mag}
    M_{\rm \lambda,MW} = M_{\rm \lambda,ref} - 2.5 \log{\left(\sum_i^{N_{\rm MAAP}}{10^{-0.4(M_{\lambda,i}-M_{\rm \lambda,ref})}}\right)},
\end{equation}
where the magnitude of the ``reference'' MAAP ($M_{\rm ref}$) is simply the first one in the array and $M_{\lambda,i}$ is the magnitude of each of the $N_{\rm MAAP}$ MAAPs. This combination is repeated separately for the dust-free (\S\ref{sec:maap_mags}) and extincted (\S\ref{sec:reddening}) magnitudes. We use the reddened colors for comparison to observations (e.g., \S\ref{sec:mw_today}) and dust-free colors for comparison to (dust-free) simulations (\S\ref{sec:sim_analogs}).

Uncertainties on the magnitudes and resultant colors are estimated in a Monte-Carlo fashion, using the 25\% mass uncertainties adopted in \S\ref{sec:maaps} and the final masses (\S\ref{sec:extrapolation}) and SSP magnitudes (\S\ref{sec:maap_mags}) associated with each MAAP. The uncertainty for a single MAAP total magnitude is $\sim$0.27~mag, and the uncertainty on the combination of many MAAPs (Eq.~\ref{eqn:total_mag}) depends primarily on the range of MAAP magnitudes (increases with range of luminosities) and the number of MAAPs being combined (decreases with larger $N_{\rm MAAP}$). For representative values of all of these quantities, we find consistent integrated $M_{\rm \lambda,MW}$ uncertainties of $\sim$0.05~mag, so we adopt 0.07~mag as a representative uncertainty on integrated MW color (e.g., \S\ref{sec:mw_today}). 

\subsection{Accounting for Stellar Evolution in MAAP Masses}
\label{sec:different_masses}
To aid comparison with literature and to use at various points later in this work, we account for stellar evolution to convert the {\it initial} MAAP stellar masses ($M_{\rm init}$; \S\ref{sec:maaps}) to i) ``aged'' stellar masses, which are the initial values minus the mass loss due to winds/SNe (but including stellar remnants), and ii) ``luminous'' stellar masses, which disregard both mass returned to the ISM and stellar remnants.

For simplicity, we turn again to the PARSEC isochrones and their \texttt{int\_IMF} values to provide the masses and numbers of stars no longer contributing to the light, as a function of MAAP age. To calculate remnant mass, we use the initial-final mass relations of \citet[][their Eqns.~1--4]{Cunningham_2024_WDifmr} for $M\le7$~M$_\odot$ and of \citet[][their Appendix~C]{Spera_2015_remnantmasses} for $M>7$~M$_\odot$, though we note that adopting other published relations do not affect the results. We find that metallicity plays almost no role in these fractional mass calculations over the range of MAAP metallicities considered here, so we only consider population age. The other change we make is to the upper end of the assumed IMF; the PARSEC isochrones include stars up to 350~M$_\odot$, which produces unrealistically high mass loss in young populations. We cap the IMF at $\sim$50~M$_\odot$, in practice by shifting the mass loss to exceed zero only at $\tau \ge 10$~Myr, the youngest MAAP considered in this paper. Again, the exact choice of mass limit or minimum mass-less age does not affect any of the results.

The MAAP ``luminous'' stellar masses, $M_{\rm \ell}$, are then the initial MAAP masses  minus the fully-evolved stars.  The MAAP ``aged'' stellar masses are the initial MAAP masses minus the fully-evolved stars, plus their remnant masses: $M_* = M_{\ell+r}$. Throughout the rest of this paper, we specify which mass is used in a given plot or analysis, but we emphasize that none of our conclusions depend on a particular mass set.

\section{The Integrated Milky Way Today}
\label{sec:mw_today}

\subsection{Mass and Colors}

\begin{figure*}[!hptb]
    \centering
    \includegraphics[width=\textwidth]{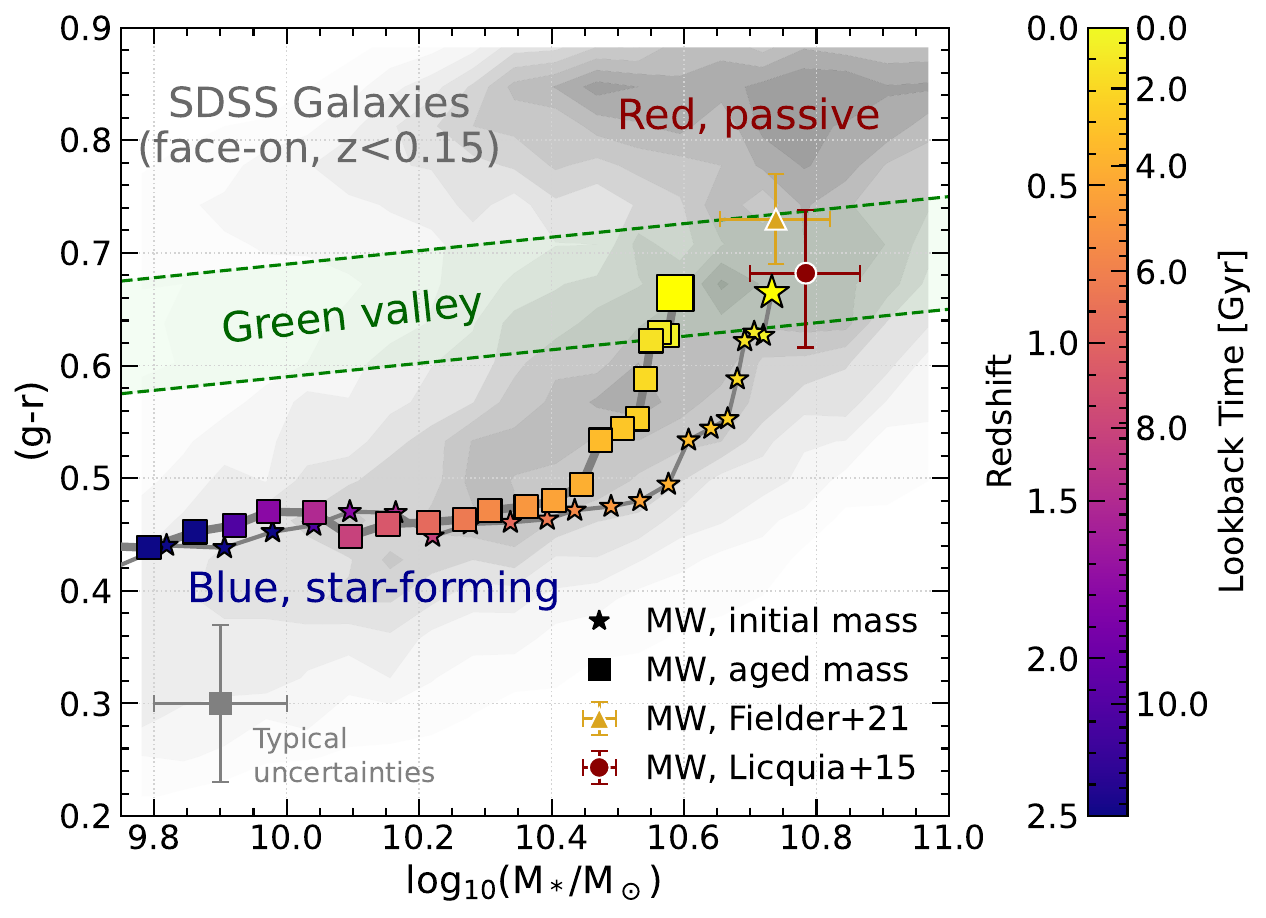}
    \caption{Comparison between the stellar masses and $(g-r)$ color for the present day MW as derived in this work (\S\ref{sec:mw_today}; large yellow square and star), in \citet[][gold triangle]{Fielder_2021_mwSED}, and in \citet[][maroon circle]{Licquia_2015_MWcolorL}. All estimates are comfortably within the ``green valley'' region indicated by the green band \citep[as defined in][]{Mendel_2013_galquenching}. The background gray contours indicate the distribution of near-face-on SDSS galaxies with $z<0.15$. The smaller colored squares and stars indicate the stellar mass and $(g-r)$ color inferred at different lookback times, as described in \S\ref{sec:time_machine}.
    }
    \label{fig:sdss_lit_compare}
\end{figure*}

Figure~\ref{fig:sdss_lit_compare} shows our results for the present-day MW stellar mass ($M_*=M_{\ell+r}$, with reddening; \S\ref{sec:reddening}) as the large yellow square in the mass~vs.~$(g-r)$ plane. (The other colored squares indicate the comparable results at higher redshifts, as presented in \S\ref{sec:time_machine}).
Shown in star symbols are the masses inferred from the summed MAAP initial stellar masses (\S\ref{sec:different_masses}).
The green band indicates the ``green valley'' region \citep{Mendez_2011_greengalaxies,Lackner_2012_bulgedisc,Mendel_2013_galquenching}, in which the MW lies, as expected \citep[e.g.,][]{Mutch_2011_greenvalley}. 

The maroon circle shows the MW stellar mass from \citet{Licquia_2015_MWmassSFR}, inferred from a meta-analysis of literature measurements, and the companion integrated $^0(g-r)$ color from \citet{Licquia_2015_MWcolorL}, derived from present-day MW mass \& star-formation analogs. The gold triangle shows the stellar mass from \citet{Fielder_2021_mwSED}, which is an update to the \citet{Licquia_2015_MWmassSFR} value, and their $^0(g-r)$ color derived using Gaussian Process Regression on the SEDs of MW structural \& star-formation analogs. Both of these mass calculations include luminous stars and stellar remnants, and thus are more directly comparable to this work's aged stellar mass indicated by the yellow square; despite being closer in value to the summed initial mass shown with the yellow star, all estimates are consistent within 1$\sigma$.

The gray contours show the mass--color distribution of near-face-on SDSS galaxies\footnote{This sample ($N\approx120,000$) was selected from the \texttt{PhotoPrimary} table view of SDSS DR17 \citep[][though the data are largely from earlier DRs]{York_2000_sdss,Abdurrouf_2021_sdssDR17}, matched to the \texttt{SpecPhoto} table for redshifts and the \texttt{stellarMassFSPSGranWideDust} table for stellar masses \citep[][\url{https://www.sdss4.org/dr17/spectro/galaxy_granada/}]{Ahn_2014_dr10}. As this was intended simply as an illustrative sample, the only selection criteria were that the $r$-band de~Vaucouleurs axis ratio $b/a$ (\texttt{deVAB\_r} from \texttt{PhotoPrimary}) was $>$0.75 and that the \texttt{SpecPhoto} redshift $\le$0.15.}, all within $z<0.15$ and with photometry k-corrected to $z=0$ using \texttt{kcorrect}\footnote{\url{https://kcorrect.readthedocs.io/}} \citep{Blanton_2007_kcorrect}. 
The density peaks corresponding to the blue star-forming main sequence and the red quiescent galaxy sequence are visible, with the green valley between them.

Other typical colors, with and without reddening, are listed in Table~\ref{tab:mw_colors}.

\begin{deluxetable}{cc|cc}
\tablecaption{Integrated Milky Way Colors (\S\ref{sec:combine_mags})}
\label{tab:mw_colors}
\tablewidth{\textwidth}
\tablehead{
\colhead{Intrinsic Color} & \colhead{Value} & \colhead{Reddened Color} & \colhead{Value}
}
\startdata
$(u-g)_0$ & 1.31 & $(u-g)$ & 1.42 \\
$(g-r)_0$ & 0.55 & $(g-r)$ & 0.66 \\
$(r-i)_0$ & 0.25 & $(r-i)$ & 0.30 \\
$(i-z)_0$ & 0.17 & $(i-z)$ & 0.21 \\
\enddata
\tablecomments{All magnitudes are in the AB system. Typical color uncertainties are $\pm$0.07 dex (\S\ref{sec:combine_mags}).}
\end{deluxetable}

\subsection{Metallicity Distributions and Gradients}
\label{sec:mdfs_and_gradients}

Metallicity distribution functions (MDFs), particularly of the solar neighborhood, have long been used to motivate and constrain models of chemical enrichment. Even in the absence of additional elements, such as $\alpha$-abundances, MDFs can reveal the importance of gas inflows/outflows and other non-closed-box galactic-scale processes \citep[e.g.,][]{Pagel_1975_snMDF}. As these processes vary in frequency and strength across the Galactic disk, knowing the distribution of heavy elements at different Galactic radii is crucial for a holistic understanding of the MW's chemical enrichment \citep[e.g.,][]{Schlesinger_2012_SEGUEMDF,Weinberg_2019_chemicalcart}.

\begin{figure*}[!pt]
    \centering
    \includegraphics[width=\textwidth]{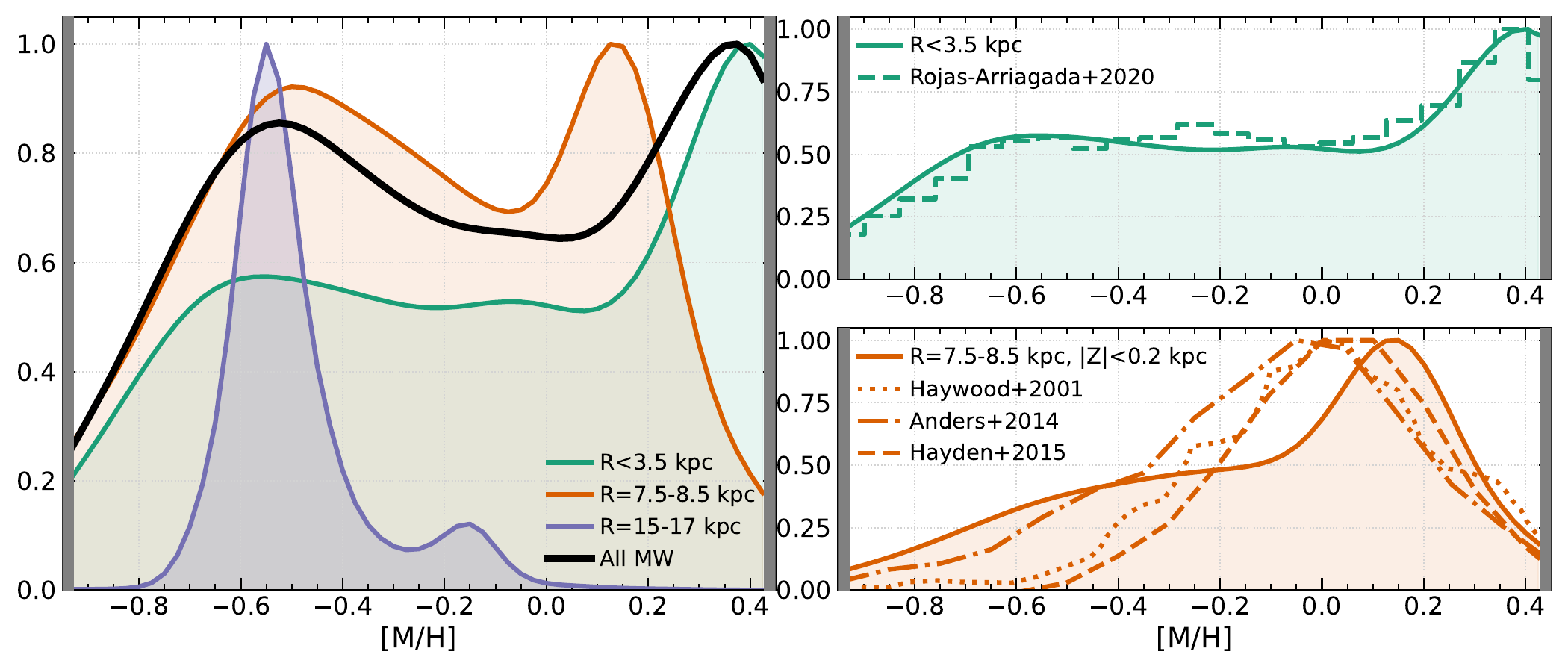}
    \caption{
    {\it Left}: The mass-weighted MDFs (scaled to unity at the peaks) derived from the MAAPs in this work (\S\ref{sec:mdfs_and_gradients}) for the inner Galaxy ($R_{\rm GC} \le 3.5$~kpc; green), the solar circle ($7.5 \le R_{\rm GC} \le 8.5$~kpc; orange), and an annulus of the outer disk ($15 \le R_{\rm GC} \le 17$~kpc; purple). These include mass at all $|Z|$ within the given radii. The total MW distribution is shown in black.
    {\it Top Right}: Comparison of the inner Galaxy MDF to that of \citet{RojasArriagada_2020_bulgeMDF}. 
    {\it Bottom Right}: Comparison of the solar circle MDF, now restricted to MAAP mass within $|Z| < 0.15$~kpc, to that of \citet{Haywood_2001_snMDF}, \citet{Anders_2014_MWchemodynamics}, and \citet{Hayden_2015_diskMDFs}. 
    See \S\ref{sec:mdfs_and_gradients} for details.
    In all panels, the vertical gray bars represent the edges of the metallicity range considered.
    }
    \label{fig:mdf}
\end{figure*}

The left-hand panel of Figure~\ref{fig:mdf} shows kernel density estimates (KDEs) of the mass-weighted MDFs inferred from the MAAP profiles for three regions in the MW: the bulge ($R_{\rm GC} \le 3.5$~kpc; green), the solar circle ($7.5 \le R_{\rm GC} \le 8.5$~kpc; orange), and an annulus of the outer disk ($15 \le R_{\rm GC} \le 17$~kpc; purple). All of these selections include the mass at all $|Z_{\rm GC}|$ within those annuli. We use the MAAPs' aged stellar mass, $M_{\ell+r}$ for weighting here, but adopting either of the other masses produces no noticeable difference.

Some clear and expected trends are visible, such as the overall decrease in mean [M/H] as $R_{\rm GC}$ increases.
The inner Galaxy's MDF has its highest peak at $\rm [M/H] \sim +0.3$, with a flattening (or a tiny bump) at $\rm [M/H] \sim -0.2$ and a smaller secondary bump at $\rm [M/H] \sim -0.6$. 
This is broadly similar to the suite of bulge metallicity components identified in the literature \citep[e.g.,][]{Ness_2016_bulgeMDF,Johnson_2022_BDBSmdf}, perhaps most strikingly so to the APOGEE-derived, selection-function corrected analysis of \citet{RojasArriagada_2020_bulgeMDF}, which is shown explicitly in comparison in the upper right-hand panel. Though both based on APOGEE data (albeit different data releases), these MDFs are derived in very different ways, so their close similarity in shape highlights the robustness of the complexity of the metallicity distribution in the inner MW.

On the other hand, the distribution representing the solar annulus is broader than typically cited for the regions near the Sun, with a primary peak at $\rm [M/H] \sim +0.15$ and a secondary peak at $\rm [M/H] \sim -0.45$. However, this includes significant mass from the thicker high-[$\alpha$/M] disk, extending out beyond $|Z_{\rm GC}| > 5$~kpc, which is responsible for a large part of the (relatively) metal-poor tail. In the lower right-hand panel, we compare the MDF inferred only from the mass within $|Z_{\rm GC}| < 0.2$~kpc to those from \citet{Haywood_2001_snMDF}, \citet{Anders_2014_MWchemodynamics}, and \citet{Hayden_2015_diskMDFs}, and we find much more similar behavior, though still with a more metal-rich primary peak (by $\sim$0.1~dex) and a heavier low-metallicity tail below $\rm [M/H] \sim -0.2$. The MAAP mass comprising this tail is still heavily skewed towards the outer edge of this $|Z_{\rm GC}|$ range, suggesting a different vertical distribution from the \citet{Haywood_2001_snMDF} and \citet{Hayden_2015_diskMDFs} comparison samples in particular; the \citet{Hayden_2015_diskMDFs} number-weighted MDF is based on APOGEE (DR12) giants but is not corrected for the survey selection function, and the \citet{Haywood_2001_snMDF} distribution is based on local low-mass dwarf stars, converted to a mass distribution using isochrones. The \citet{Anders_2014_MWchemodynamics} sample (their ``APOGEE HQ Sample'') is restricted to $d \le 1$~kpc but with very few stars at $d<0.25$~kpc.

The ``outer disk'' (purple) is dominated by low-metallicity stars (and $\sim$solar-[$\alpha$/M]) at $\rm [M/H] \sim -0.6$. Interestingly, there is a (much smaller) bump at $\rm [M/H] \sim -0.2$, which approximately corresponds to enhancements seen also in the bulge and solar neighborhood MDFs. Stars in the disk beyond $R_{\rm GC} \sim 15$~kpc have been argued to be dominated by migrators from smaller Galactic radii \citep{Lian_2022_MWmigration}, and thus their MDFs are driven by both the MDFs at the smaller radii and the likelihood of migrating from those radii (which is itself a function of time or age). Thus this small bump may reflect an enhancement of more metal-rich migrators driven simply by the greater number of stars at that metallicity at smaller $R_{\rm GC}$.

The ability to produce inferred MDFs essentially anywhere in the MW demonstrates the power and potential of this MAAP-modeling approach, supported by large surveys like APOGEE and eventually Milky Way Mapper \citep[][J.A.~Johnson, in prep]{Kollmeier_2017_sdss5,Almeida_2023_dr18} with relatively simple, recoverable selection functions\footnote{Obviously, the current MAAP fits exclude a Galactic halo structure and are insensitive to non-axisymmetric populations, such as the ``bar'' vs ``non bar'' regions of the bulge. This is a limitation of sample size that will be ameliorated with future datasets.}. The presence and strength of the $\rm [M/H] \sim -0.2$ bump in the outer disk MDF above, for example, could be traced smoothly with Galactic position and compared to predictions from different radial migration models or prescriptions, or from outside-in quenching scenarios \citep[e.g., Sec.~4.2 in][]{Lian_2022_MWmigration}. As another example, the extraction of complete, mass- or light-weighted MDFs at arbitrary points in the Galaxy would provide extremely valuable constraints on modeling the radially-resolved star formation history of the Milky Way.

\begin{figure}[!hptb]
    \centering
    \includegraphics[width=0.5\textwidth]{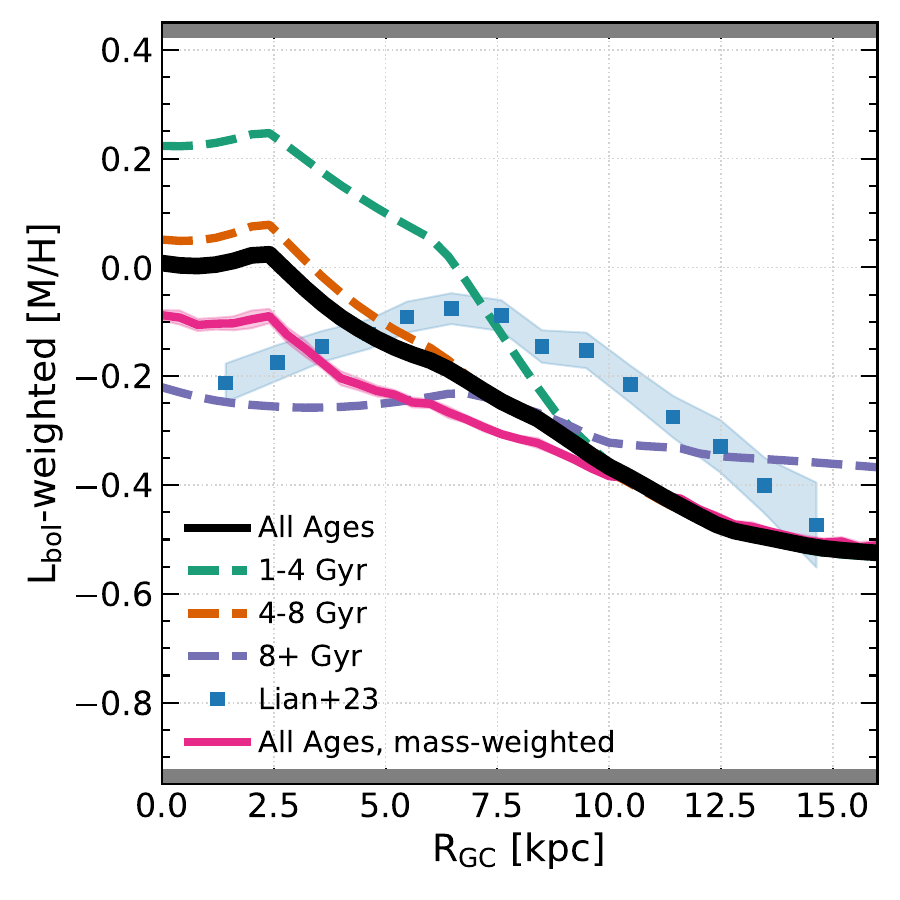}
    \caption{
    The radial metallicity gradient inferred from the MAAPs in this work (\S\ref{sec:mdfs_and_gradients}), weighted by bolometric luminosity (thick black line) and by stellar mass (hot pink line). The green, orange, and purple dashed lines indicate the $L_{\rm bol}$-weighted metallicities for MAAPs with ages 1--4~Gyr, 4--8~Gyr, and $>$8~Gyr, respectively. Shown for comparison is the light-weighted metallicity profile of \citet{Lian_2023_integratedMHprofile} in blue.
    }
    \label{fig:mh_gradient}
\end{figure}

Figure~\ref{fig:mh_gradient} shows the integrated [M/H] profile of the galaxy, weighted at each $\Delta R=0.5$~kpc radial bin by the dust-free bolometric luminosity of the MAAPs. (Due to how the reddening is computed in \S\ref{sec:reddening}, the result is the same with the extincted luminosities.) We find a steady slope of approximately \textcolor{black}{$-0.05$~dex~kpc$^{-1}$ within $R_{\rm GC} = 6-14$~kpc}, with flatter slopes ($\sim$0~dex~kpc$^{-1}$) both inside and outside that range\footnote{The small kink at $R_{\rm GC} \sim 3$~kpc is most likely due to our simple correction of the oversimplified inner density profile (\S\ref{sec:maaps}), so we do not place importance on it.}. Given the luminosity-weighted mean age of $\sim$5~Gyr at these radii, this value is consistent with other slopes derived for similarly-aged stars \citep[e.g.,][]{Anders_2023_agetrends,Lian_2023_integratedMHprofile,Vickers_2021_MHgradient} and open clusters \citep[e.g.,][]{Donor_2020_occam4,Myers_2022_occamDR17}, albeit with a range of absolute offsets among this comparison sample. Likewise, the youngest stars (represented by the $<$4~Gyr line) have a steeper slope in this same radial range, and the oldest stars (represented by the $>$8~Gyr line) have the shallowest, consistent with trends seen in both observations \citep[e.g.,][]{Anders_2023_agetrends} and simulations \citep[e.g.,][]{Graf_2024_FIREabundances}. We note that weighting by different bandpass luminosities does shift the absolute values of the [M/H] profiles, but does not strongly affect the slopes.

In Figure~\ref{fig:mh_gradient}, we also overplot the conceptually similar light-weighted metallicity profile of \citet{Lian_2023_integratedMHprofile}, which notably shows a break at $R_{\rm GC} \sim 7$~kpc. We observe a small inflection, a flattening, of the gradient inwards of this radius, but not a clear reversal of slope. In that work, the turnover is attributed to the domination of the inner Galaxy light by the oldest populations, which have the lowest [M/H] in these regions. In contrast, our 4--8~Gyr MAAPs carry a larger fraction of the inner Galaxy luminosity, which produces a flat gradient at roughly solar metallicity inside the 6~kpc break radius. Notably, the equivalent (aged) \emph{stellar mass}-weighted [M/H] profile is also flat in the inner Galaxy, but at a lower value ($\rm [M/H] \sim -0.1$) than the bolometric luminosity-weighted one ($\rm [M/H] \sim 0$).

The fact that we see such complexity in the profiles, regardless of exactly where the breaks and turnovers are, highlights the need for caution when comparing gradients extracted from different observational tracers in the MW, in other galaxies, and in simulations.

\subsection{Spatially Resolved Star Formation Histories}
\label{sec:sfhs}

As with MDFs, star formation histories (SFHs) or historical star formation rates are important psuedo-observables of galaxy evolution for chemical evolution modeling and comparison to simulations. From the observational perspective, one important question is how applicable constraints on the {\it local} (i.e., solar neighborhood) SFH are for understanding that of the MW in general.

\begin{figure*}[!hptb]
    \centering
    \includegraphics[width=\textwidth]{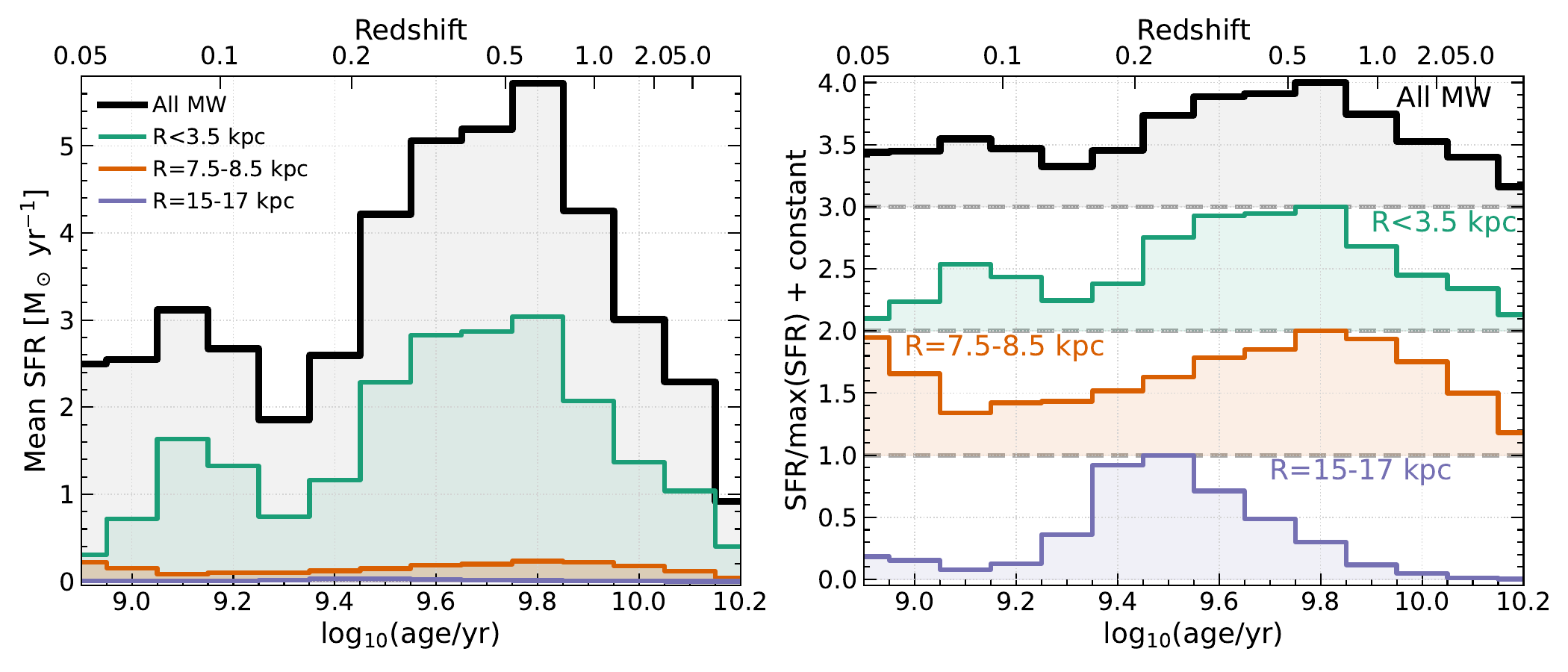}
    \caption{
    Mean star formation rate (SFR) as a function of time (\S\ref{sec:sfhs}), inferred from the distribution of mass as a function of age in the inner Galaxy ($R_{\rm GC} \le 3.5$~kpc; green), the solar circle ($7.5 \le R_{\rm GC} \le 8.5$~kpc; orange), and an annulus of the outer disk ($15 \le R_{\rm GC} \le 17$~kpc; purple). These include mass at all $|Z|$ within the given radii. The ``total'' MW distribution is shown in black.
    {\it Left}: The absolute SFR in units of M$_\odot$ per year. 
    {\it Right}: The SFR of each location, scaled to unity at its maximum value and offset vertically for clarity. 
    }
    \label{fig:sfr}
\end{figure*}

The left-hand panel of Figure~\ref{fig:sfr} shows the mean star formation rate (SFR) as a function of stellar age inferred from the MAAPs (using $M_{\rm init}$) in the same broad spatial regions used in \S\ref{sec:mdfs_and_gradients}: the bulge ($R_{\rm GC} \le 3.5$~kpc; green), the solar circle ($7.5 \le R_{\rm GC} \le 8.5$~kpc; orange), and an annulus of the outer disk ($15 \le R_{\rm GC} \le 17$~kpc; purple). The right-hand panel shows the same information, but with the SFR(t) values normalized to unity at each region's peak (and vertically offset) to highlight differences in the {\it relative} rates over time, rather than absolute values. We emphasize that these are not true ``star formation histories'' of a particular location in the MW, but simply the historical SFRs needed to produce the stars that lie at those locations in the MW {\it today}. 

The ``all MW'' line --- as a reminder, representing stars with $\rm -0.95 \le [M/H] \le +0.45$ --- has two clear peaks, a very broad one at $\log{\rm \tau/yr} \sim 9.7$ (corresponding to $\rm \tau \sim 5$~Gyr and $z \sim 0.5$) that has a long tail to earlier times, and a narrower one at $\log{\rm \tau/yr} \sim 9.2$ ($\rm \tau \sim 1.6$~Gyr, $z \sim 0.1$). Since this distribution represents nearly all stars in the MW today, independently of radial migration or other processes that can move stars around within the Galaxy, it is the closest (of all the lines in the plot) to a ``true'' SFH for its associated region.

As with the MDFs, the {\it shape} of the full MW's pattern is most closely matched by that of the inner Galaxy. Both the solar annulus and the outer disk regions have far lower absolute SFRs, but also different shapes over time, as more easily seen in the right-hand panel of Figure~\ref{fig:sfr}. The solar annulus shares the bulge's steep increase in star formation at early times, but then levels off to a roughly constant rate, before a small increase at $\log{\rm \tau/yr} \sim 9.2$ and then a decrease to today's low-but-nonzero rate. This picture is qualitatively consistent with that portrayed by age--metallicity--abundance distributions in different parts of the Galaxy \citep[e.g.,][]{Isern_2019_WDsfh,Lian_2020_bulgeSFH,Spitoni_2023_3infall,Nataf_2024_subgiants,Gallart_2024_GaiaSFH}: an early, centrally-concentrated burst of star formation, followed by a quenching primarily in the inner Galaxy while star formation continued at a lower level in the disk, and then smaller reinvigorations $\sim$8~Gyr and (especially near the solar annulus) $\sim$2--3 Gyr ago. 

The outer disk's present-day stars have a very different history: a dearth of both very old stars and very young stars, with a single-but-broad peak in the SFR centered at $\log{\rm \tau/yr} \sim 9.5$ ($\rm \tau \sim 3.2$~Gyr) sufficient to explain the present-day stellar population. As this region of the disk is likely even more heavily impacted by radial migration and satellite perturbations than the other examples shown, this peak likely represents some optimal combination of birth radius $R_{\rm birth}$, the star formation history at $R_{\rm birth}$, satellite interaction history \citep[e.g., with the Sagittarius dwarf galaxy;][]{RuizLara_2020_Sgr}, and migration efficiency to populate these far reaches of the disk.

\section{Running the Clock Backwards}
\label{sec:time_machine}

\subsection{Calculating MW Properties At Earlier Times}
\label{sec:time_machine_method}
To calculate the MW's stellar mass and integrated colors at earlier times, in summary, we 1) ``de-age'' all MAAPs by a particular time step, 
2) sum the masses and recompute new absolute magnitudes (as in \S\ref{sec:maap_mags}) for the remaining MAAPs, 3) compute a new set of extincted magnitudes (as in \S\ref{sec:reddening}), and then 4) combine the raw and extincted MAAP magnitudes (separately) for raw and reddened integrated values (as in \S\ref{sec:mw_today}).

In greater detail --- we define a set of ``lookback ages'' ($\vec{t}$, with a corresponding set of redshifts $\vec{z}$) \textcolor{black}{from 0.1--13.3~Gyr ($z=10^{-4}-9.8$), with a linear spacing of 0.15~Gyr}. At each lookback age step $t_i$, we subtract that value from each MAAP's age to produce a new age $\tau_i$, and remove all MAAPs with $\tau_i<0.01$~Gyr (i.e., those MAAPs with an age less than 10~Myr at that lookback age or not yet formed). After computing the mass loss due to stellar evolution as in \S\ref{sec:different_masses}, the total initial, aged, and luminous masses of the remaining MAAPs are summed and saved as the stellar masses of the MW at that age, $m_{\rm MW}(t_i)$.

Given the new MAAP age $\tau_i$ at lookback age step $t_i$, and assuming that the MAAP metallicity remains constant over time, we match each MAAP to the appropriate PARSEC stellar population (as in \S\ref{sec:maap_mags}, Eq.~\ref{eqn:maap_mags}) to compute updated {\it ugriz} magnitudes. The updated MAAP magnitudes are then combined as in \S\ref{sec:mw_today} for the integrated magnitudes of the MW at that age, $M_{\rm \lambda,MW}(t_i)$.

Extincted magnitudes are also computed at each age step, using the same dust disk model as at $z=0$ (\S\ref{sec:reddening}). In reality, since the radial mass fractions of each MAAP are expected to change over time, due to radial migration and other processes, {\it and} the dust disk itself evolves over time, the MAAP-specific extinction should also change. Accounting for such a detailed ISM evolution (including changes in the dust properties and $A_\lambda/A_V$ behavior) is beyond the scope of this work.
Qualitatively, if the dust disk remains static, we may expect a slight shift to redder colors at higher redshifts, as today's oldest and reddest populations would have been even more centrally concentrated and thus more strongly extinguished earlier in their lifetimes. However, given that the difference between the extincted and completely extinction-free $(g-r)$ colors of the inferred $z=0$ MW (\S\ref{sec:mw_today}) is only $\sim$0.1~dex, we expect that this shift would be very small, and even smaller if the dust disk grows with the stellar disk.

The end result of this process is a set of integrated MW masses, integrated dust-free colors, and integrated reddened colors at a range of redshifts. We use a cubic spline to interpolate masses and colors as functions of redshift at the $z_{\rm select}$ and $z_{\rm track}$ values used in \S\ref{sec:sim_analogs}. \textcolor{black}{Subsets of these quantities are given in Table~\ref{tab:mw_over_time} and shown in Figure~\ref{fig:sdss_lit_compare}, and used in the analysis in later sections.}

\begingroup
\renewcommand{\arraystretch}{1.3} 
\begin{deluxetable*}{cc|ccccc}
\tablecaption{Integrated Milky Way Properties Over Time (\S\ref{sec:time_machine_method})}
\label{tab:mw_over_time}
\tablewidth{\textwidth}
\tablehead{
\colhead{\multirow{3}{*}{$z$}} & \colhead{\multirow{3}{*}{age [Gyr]}} & 
\colhead{$\log(M/M_\odot)^a$} & 
\colhead{\multirow{2}{*}{$(g-r)_0$}} & \colhead{\multirow{2}{*}{$(g-r)$}} & \colhead{\multirow{3}{*}{SFR [M$_\odot$~yr$^{-1}$]}} & \colhead{\multirow{3}{*}{Examples$^b$}} \\
& & ($M_{\rm init}$ -- $M_{\ell+r}$ -- $M_{\ell}$) & & & & \\
& & ($\pm$0.1) & ($\pm$0.07) & ($\pm$0.07) & & 
}
\startdata
0.00 & 0.00 & 10.73 -- 10.59 -- 10.51 & 0.55 & 0.66 & 1.9 & \multirow{4}{*}{\includegraphics[trim=0cm 0cm 0cm 0cm, clip, width=1.5cm]{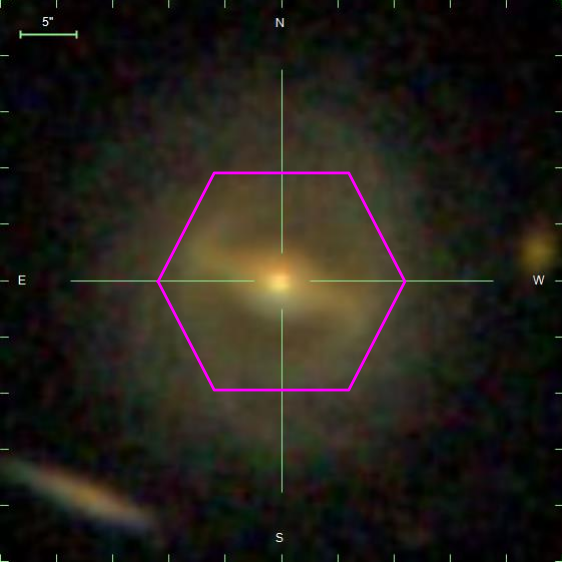}\hspace{0.1cm}\includegraphics[trim=0cm 0cm 0cm 0cm, clip, width=1.5cm]{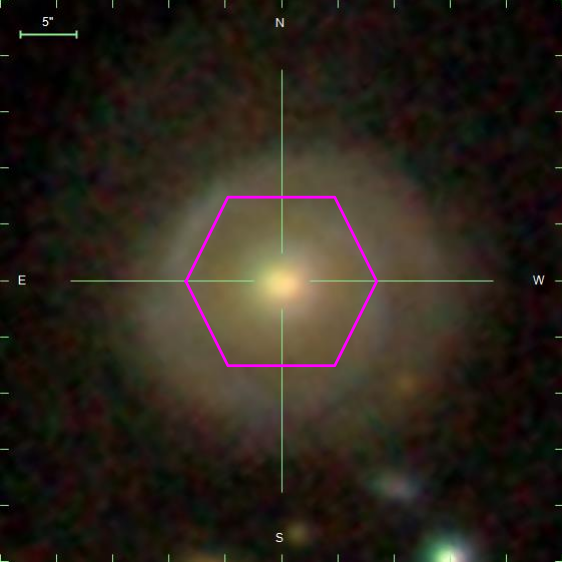}} \\
0.05 & 0.70 & 10.72 -- 10.57 -- 10.50 & 0.48 & 0.62 & 2.5 & \\
0.10 & 1.35 & 10.70 -- 10.56 -- 10.48 & 0.51 & 0.64 & 1.9 & \\
0.20 & 2.51 & 10.68 -- 10.54 -- 10.47 & 0.43 & 0.57 & 2.7 & \\
0.50 & 5.20 & 10.55 -- 10.42 -- 10.35 & 0.31 & 0.49 & 5.4 & \multirow{2}{*}{\includegraphics[trim=0cm 0cm 0cm 0cm, clip, width=1.2cm]{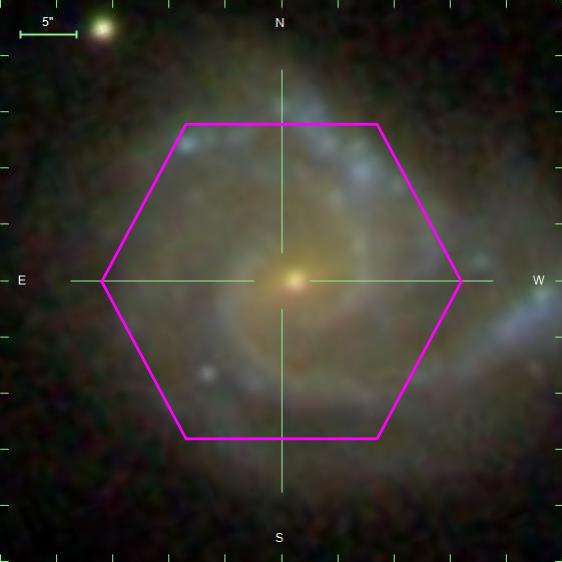}} \\
1.00 & 7.94 & 10.31 -- 10.19 -- 10.13 & 0.29 & 0.48 & 4.1 & \\
1.50 & 9.52 & 10.17 -- 10.05 -- 9.99 & 0.27 & 0.47 & 3.8 & \multirow{2}{*}{\includegraphics[trim=0cm 0cm 0cm 0cm, clip, width=1.2cm]{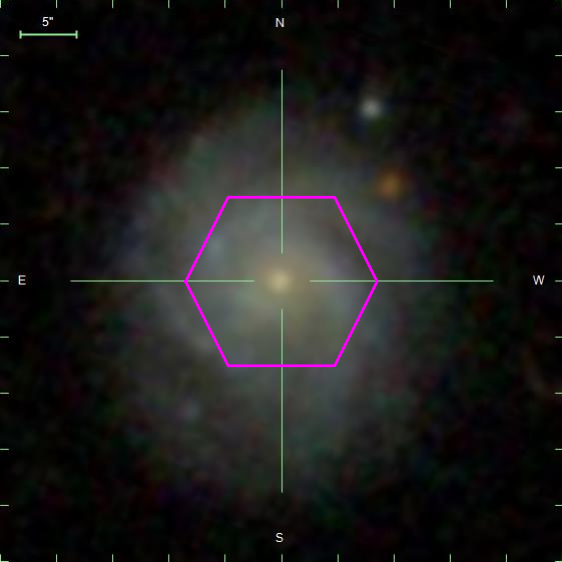}} \\
2.00 & 10.51 & 10.06 -- 9.94 -- 9.89 & 0.28 & 0.46 & 2.4 & \\
\enddata
\tablecomments{$^a$\,Stellar mass types defined in \S\ref{sec:different_masses}. $^b$\,Selected from MaNGA \citep{Bundy_2015_manga} using galaxy stellar colors \& masses (as in \S\ref{sec:tng_description} below) from \citet{Sanchez_2022_pipe3d}.}
\end{deluxetable*}
\endgroup

\subsection{The Integrated Milky Way Over Time}
\label{sec:time_machine_results}

In addition to the mass and $(g-r)$ color of the present-day MW and $z<0.15$ SDSS galaxies, Figure~\ref{fig:sdss_lit_compare} shows the evolution of these properties for the MW as inferred from the ``youthening'' process described in \S\ref{sec:time_machine_method}. We can see that the MW has been in the green valley for approximately 1.2~Gyr, since $z\sim 0.08$. This is necessarily a very coarse approximation, given the fuzziness of the green valley boundary definitions and the integrated color uncertainties, but it is reassuring that the measurements do not suggest that the MW has been in this transitional phase for a long time \citep[though see also][]{Schawinski_2014_greenvalley}. We also note that though it occupies the relatively underpopulated green valley, the MW appears to lie near the most densely populated mass range within that valley.

Looking back further in time, at least until $z \sim 1$, the MW's path in this plane very closely {\it parallels} the low-$z$ blue star-forming main sequence of galaxies, with a similar slope but slightly either bluer at fixed mass or more massive at fixed color. We address possible explanations for this offset in terms of star formation rate in \S\ref{sec:sim_analogs}. At higher redshifts, the MW's track plateaus at a roughly constant $(g-r) \sim 0.47$, consistent with young populations of $\tau \lesssim 2$~Gyr, even younger when the effects of dust are considered.

\begin{figure}[!hptb]
    \centering
    \includegraphics[width=0.5\textwidth]{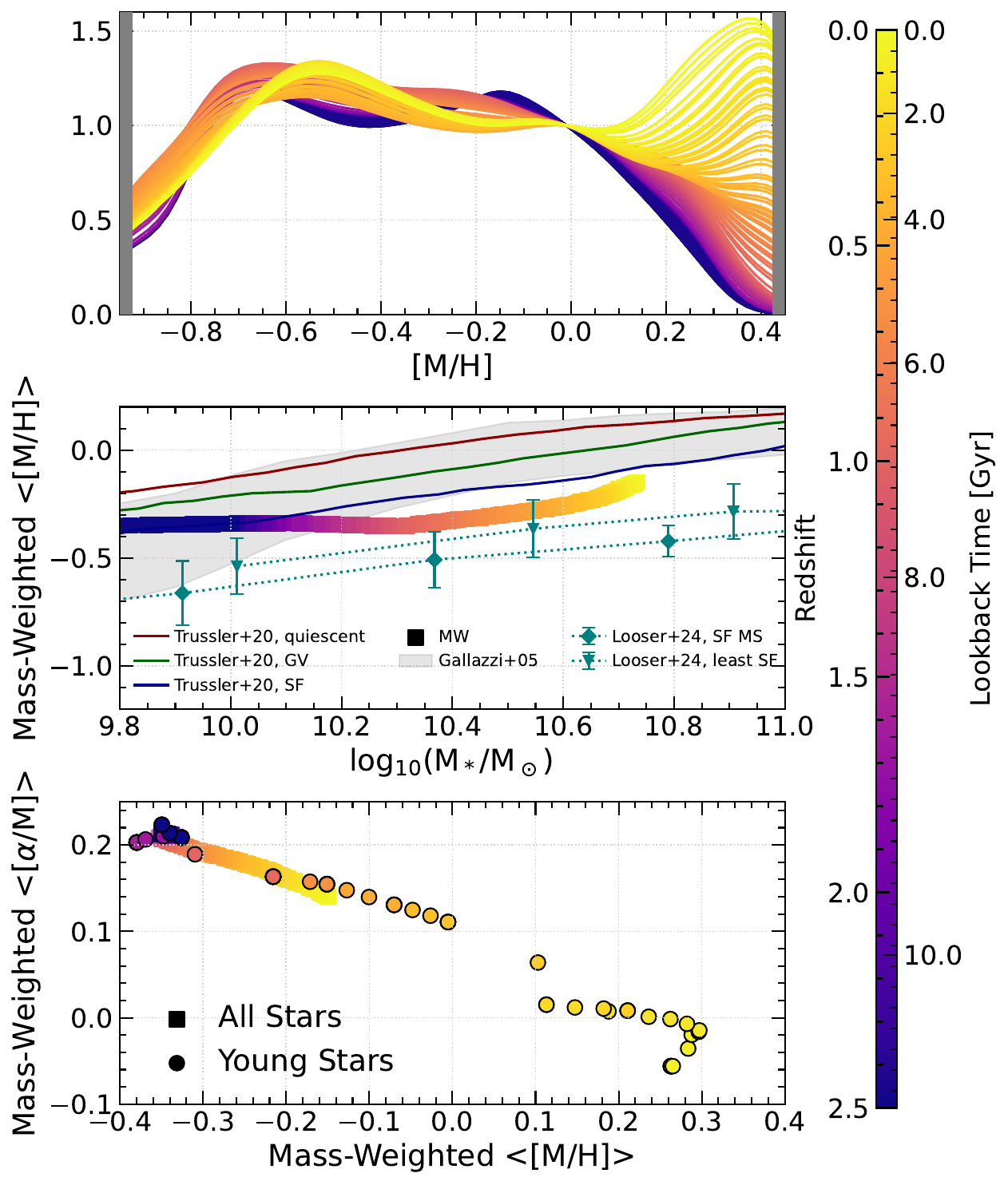} 
    \caption{Integrated, stellar-mass-weighted metallicities over time (\S\ref{sec:time_machine_results}).
    {\it Top:} A gaussian KDE of the integrated, mass-weighted MDF at different redshifts. All have been scaled to pass through unity at solar metallicity, to highlight the shape change rather than absolute values.
    {\it Middle:} The stellar mass--stellar metallicity plane, showing our inferred MW points at different redshifts (colored squares) compared to various galaxy samples.
    {\it Bottom:} Similar to the middle panel, but for mass-weighted [M/H] and [$\alpha$/M]. The colored squares are identical to the panel above, and the circles show values from stars with $\tau < 200$~Myr.}
    \label{fig:mdf_vs_z}
\end{figure}

The top panel of Figure~\ref{fig:mdf_vs_z} shows the shape evolution of the MW's integrated, mass-weighted MDF over time. Each MDF is a gaussian KDE representation using a 0.2~dex kernel, weighted by the aged stellar mass at each [M/H] at each lookback time (indicated by the line color) and scaled to pass through unity at solar metallicity. The present-day $z=0$ (yellow) line is thus identical to the ``All MW'' MDF in Figure~\ref{fig:mdf}. We can see that as we look further back in time, in general, the MDF becomes more heavily weighted towards lower metallicities, but the evolution is not monotonic with time at subsolar metallicities. This is due to the multiple ``bursty'' star formation events experienced at different times by different parts of the Galaxy that had been enriched to different levels (\S\ref{sec:sfhs} and Figure~\ref{fig:sfr}). 

In the middle panel of Figure~\ref{fig:mdf_vs_z}, we show the relationship over time between the MW's stellar mass (including remnants) and mass-weighted mean stellar metallicity. This is akin to the stellar mass--metallicity relation \citep[SMZR; e.g.,][]{Gallazzi_2005_smzr,Leethochawalit_2018_smzr}, itself a variant of the well-studied mass--(gas) metallicity relation \citep{Tremonti_2004_mzr}. For comparison are included a few examples of extragalactic SMZR relations. The gray shaded region is the sample of quiescent$+$star-forming galaxies from \citet{Gallazzi_2005_smzr}, with the majority of the sample at $z<0.1$. The teal points are from \citet[][]{Looser_2024_smzr}, who analyzed MaNGA galaxies with different levels of star formation relative to the star-forming main sequence; the more metal-rich sequence (triangles) represents the most quiescent galaxies, while the diamonds trace the bin in which the present-day MW lies\footnote{Assuming SFR$_{\rm MW} = 1-2$~M$_\odot$~yr$^{-1}$ and $M_{\rm *,MW} = 5.6 \times 10^{10}$~M$_\odot$; see Eqs.~1--2 and Fig.~3 of \citet{Looser_2024_smzr}.}. The red, green, and blue lines indicate the SMZR for quiescent, green valley, and star-forming galaxies, respectively, from the $z<0.085$ sample of \citet[][]{Trussler_2020_quenching}. Our inferred MW values lie well within the range of metallicities spanned by these studies at a given mass. 

In the bottom panel of Figure~\ref{fig:mdf_vs_z}, we show the integrated, mass-weighted mean [M/H] and [$\alpha$/M] over the same span of redshift. In squares are the values for all stars (identical to the data in the middle panel), and in circles are the values when only populations with $\tau < 200$~Myr are considered. As this timeframe (very roughly) encompasses the main-sequence lifetime of most OB stars, these measurements may be considered approximations to the metallicities and abundances inferred from extragalactic HII regions. They show a larger range of values than the all-star averages, and are strongly reminiscent of the bimodal $\alpha$-element distribution in the MW, particularly in the inner Galaxy \citep[e.g.,][]{Hayden_2015_diskMDFs,Zasowski_2019_bulgeabundances,Imig_2023_MWdisks}. Interestingly, recent work has shown similar complex sequences in star-forming galaxies in the analogous $\rm [O/Ar]-[Ar/H]$ plane \citep[e.g.,][]{Bhattacharya_2025_SFGalpha,Esteban_2025_SFGalpha}.

A detailed analysis of the differences between such extragalactic and MW-derived abundances, and how the MW's evolution might in interpreted in terms of how its gradual quenching has been impacted by accretion and outflows \citep[e.g.,][]{Peng_2015_quenching,Trussler_2020_quenching,Looser_2024_smzr}, is beyond the scope of this paper. But we include this figure as yet another example of how these time-resolved measurements of the young, integrated MW may be used as a powerful diagnostic to complement the statistical, single-snapshot nature of extragalactic samples.

\section{Effect of Gaia-Enceladus and Sagittarius Populations}
\label{sec:mergers}

\begin{figure*}[!hptb]
    \centering
    \includegraphics[trim={0in 0in 0in 0in},width=\textwidth]{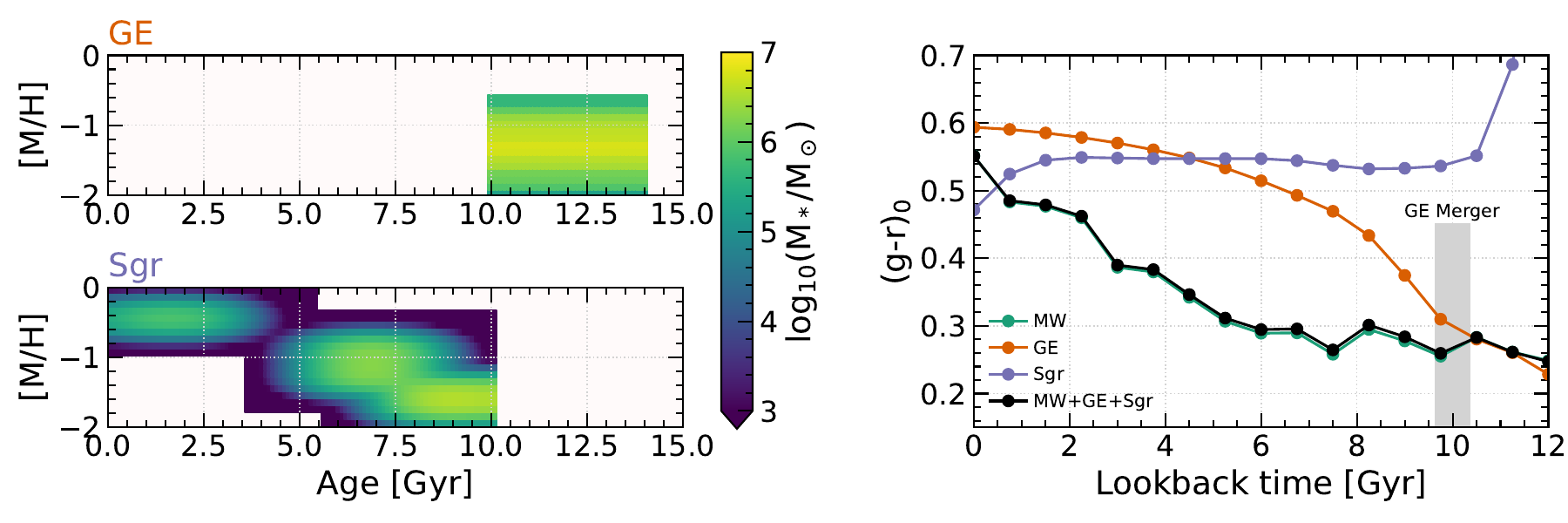} 
    \caption{
    Color evolution of the MW with and without including stars belonging to Gaia-Enceladus and Sagittarius (\S\ref{sec:mergers}).
    {\it Left}: MAAP masses in the age--metallicity plane (akin to Figure~\ref{fig:maap_masses}) for GE (top) and Sgr (bottom).
    {\it Right}: Evolution of the integrated $(g-r)_0$ color for the MAAPs associated with the MW only (green), GE (orange), Sgr (purple), and all three systems together (black).
    }
    \label{fig:merger_colors}
\end{figure*}

The MAAPs analyzed thus far span the vast majority of stars in the MW's disk and bar/bulge, which comprise nearly all the stellar content of the Galaxy. The remaining stars reside in the halo, which is dominated by the remnants of past galactic mergers \citep[e.g.,][]{Naidu_2020_halostruct,Horta_2023_halostruct}. Among the largest of these are i) Gaia-Enceladus \citep[GE;][]{Helmi_2018_ges,Belokurov_2018_GES}, whose stars are spatially well-mixed with {\it in situ} stars but can be identified via their orbital properties and/or chemistry \citep[e.g.,][]{Feuillet_2021_gesseq,Carrillo_2024_ges}, and ii) the Sagittarius dwarf galaxy \citep[Sgr; e.g.,][]{Ibata_94_SgrdSph,Majewski_03_SgrdSph}, which maintains a distinct galactic core connected to chemically-stratified leading and trailing tidal tails that together wrap multiple times around the Milky Way \citep[e.g.,][]{Belokurov_2006_FoS,Hayes_2020_sgrstream,Vasiliev_2021_sgrstream}.

We were curious as to whether the stellar populations of these two systems, if included in the MW's integrated properties, have a measurable effect on those properties, especially at early times when the MW was smaller. Stars formed in the MW itself as a result of the past (in the case of GE) and ongoing (in the case of Sgr) merger events are generally already represented in our existing MAAPs, so we focus on the stars belonging to the smaller galaxies themselves. The GE stars are not included in our MAAPs because of their low metallicity, and the Sgr stars are not well-represented because of their preferential location in the stellar halo. Thus, to incorporate them into the integrated Galaxy measurements, we essentially created two new suites of MAAPs, representing the distribution of stellar mass in the $\rm \tau-[M/H]$ plane, based solely on literature studies. 

For GE, we assumed a total stellar mass of $1.5 \times 10^9$~M$_\odot$ and a flat age distribution between \textcolor{black}{$\tau = 10-14$~Gyr} \citep[e.g.,][]{Myeong_2019_GESSequoia,Feuillet_2020_gesMDF,Limberg_2022_ges}. We compiled several measurements of the MDF \citep[from][]{Helmi_2018_ges,Myeong_2019_GESSequoia,Mackereth_2019_accretedhalo,Feuillet_2020_gesMDF,Limberg_2022_ges}, and since these largely agree on the overall shape, we simply took the median stellar fraction, across all of the studies, in each 0.1~dex bin. See the top left panel of Figure~\ref{fig:merger_colors}. 

For Sgr, we assumed a total stellar mass (core$+$arms/tails) of \textcolor{black}{$6 \times 10^8$~M$_\odot$} \citep[e.g.,][]{NiedersteOstholt_2010_sgrmass,NiedersteOstholt_2012_sgrmass}. The age distribution of Sgr stars is more complex than the $\sim$uniformly old GE stars, and the MDF is both wider and more structured. We collected a large set of literature measurements of Sgr's star-formation history and metallicity distribution, focusing on works that characterize the ``bulk'' stellar populations of the core and arms (e.g., not limited to the globular clusters, RRL, or other special populations): \citet{Siegel_2007_sgrcore,deBoer_2015_sgrstreamSFH,Chou_07_SgrStreamFeH,Yang_2019_sgrstream,Layden_2000_sgrcoreSFH,Bellazzini_1999_sgrcoreFEH,Bellazzini_2006_sgrcoreage}. We then roughly synthesized these into three ``MAAPs'', each smoothed over a few bins in $\tau$ and [M/H], as shown in the lower left panel of Figure~\ref{fig:merger_colors}. 

With both sets of dwarf galaxy populations, we then repeated all of the steps outlined in \S\ref{sec:maap_mags}, \S\ref{sec:combine_mags}, and \S\ref{sec:time_machine_method} to generate time-resolved galaxy masses and colors. The right-hand panel of Figure~\ref{fig:merger_colors} shows the results of including these populations, weighted by stellar mass, in the total MW integrated color. Essentially, we find that the addition of either GE or Sgr, or both together, has negligible impact on the MW's color over time, even at earlier times when the MW's stellar mass was closer to that of the GE progenitor. The difference between the MW-only and MW+GE+Sgr $(g-r)_0$ color does increase at earlier times but is never larger than 5~mmag, indistinguishable from zero given our MAAP color uncertainties.

\section{Simulated Analogs Across Cosmic Time}
\label{sec:sim_analogs}

\subsection{TNG50 Subhaloes}
\label{sec:tng_description}
In this section, we compare our inferred past and present-day integrated MW properties with those of subhaloes from the TNG50 simulations of the IllustrisTNG project\footnote{\url{https://www.tng-project.org/}} \citep[specifically, the TNG50-1 suite;][]{Nelson_2019_tngDR,Nelson_2019_tng50outflows,Pillepich_2019_tng50discs}. The TNG50 simulations are the highest-resolution realizations of the cosmologically-scaled, magnetohydrodynamical IllustrisTNG project. They follow dark matter, black holes, magnetic fields, gas, and stars within a (51.7~comoving~Mpc)$^3$ cube, with a baryonic mass particle resolution of $8.5 \times 10^4$~M$_\odot$. Where needed, we adopt the same cosmological parameters as in the initial conditions for the simulations \citep{Pillepich_2019_tng50discs,PlanckCollab_2016_cosmology}, specifically $H_0 = 100h$~km~s$^{-1}$~Mpc$^{-1}$ and $h=0.6774$.

In summary, we select ``analog'' galaxies whose stellar mass and dust-free $(g-r)_0$ color matched that of our MW at the same redshift\footnote{The MW's aged stellar mass ($M_{\ell+r}$; \S\ref{sec:different_masses}) is used for this selection, but we emphasize that the qualitative conclusions do not change if either of the other mass estimates are used. We also acknowledge there are many more parameters that could be used to define meaningful analogs, such as morphology and environment, but these are either unavailable in the simulations or significantly more complex to extract, and are deferred to future work.}. We then track the stellar mass, color, and other simulated properties forwards and backwards in time at a range of redshifts, to explore how the mass--color evolutionary paths of galaxies that were similar to the MW at different points in time compare to the MW's path. This process is described in greater detail in this subsection, and the results are shown in \S\ref{sec:tng_analog_paths}.

We first define a set of redshifts at which to select simulated analogs: \textcolor{black}{$z_{\rm select} = [0, 0.1, 0.5, 1, 1.5, 2.3]$, corresponding to TNG50 snapshots [99, 91, 67, 50, 40, 30]}, and a larger set of redshifts through which to track those analogs: \textcolor{black}{$z_{\rm track} = [0, 0.05, 0.1, 0.2, 0.3, 0.4, 0.5, 1, 1.5, 2, 2.3]$, corresponding to TNG50 snapshots [99, 95, 91, 84, 78, 72, 67, 50, 40, 33, 30]}. We downloaded the group catalogs for the ``select'' snapshots\footnote{\url{https://www.tng-project.org/data/downloads/TNG50-1/} Stellar masses are in field \texttt{SubhaloMassInRadType[4]} in the group catalogs, $g$-band photometry is in \texttt{SubhaloStellarPhotometrics[4]}, and $r$-band is \texttt{SubhaloStellarPhotometrics[5]}.}. In the catalog at each snapshot, we select all subhaloes with stellar masses within \textcolor{black}{0.2~dex} of the inferred MW's mass at that redshift, and with dust-free $(g-r)_0$ colors within \textcolor{black}{0.15~dex} of the corresponding MW's color (i.e., within $\sim$2$\sigma$ of each measurement; \S\ref{sec:combine_mags}). 

Then, for up to 100 of the analogs at each $z_{\rm select}$ snapshot, we query the corresponding \texttt{lhalotree} file\footnote{\url{http://www.tng-project.org/api/TNG50-1/snapshots/[snapshotID]/subhhalos/[subhaloID]}} and extract the mass, color, and other properties backwards and forwards in time, at each of the $z_{\rm track}$ redshifts. For earlier snapshots, this information is contained in the \texttt{lhalotree} file itself; for later snapshots, we follow the \texttt{sublink\_descendant} fields in the linked subhalo files to extract the same information. 

\subsection{Simulated MW Analogs Over Time}
\label{sec:tng_analog_paths}

\begin{figure*}[!ht]
    \centering
    \includegraphics[width=\textwidth]{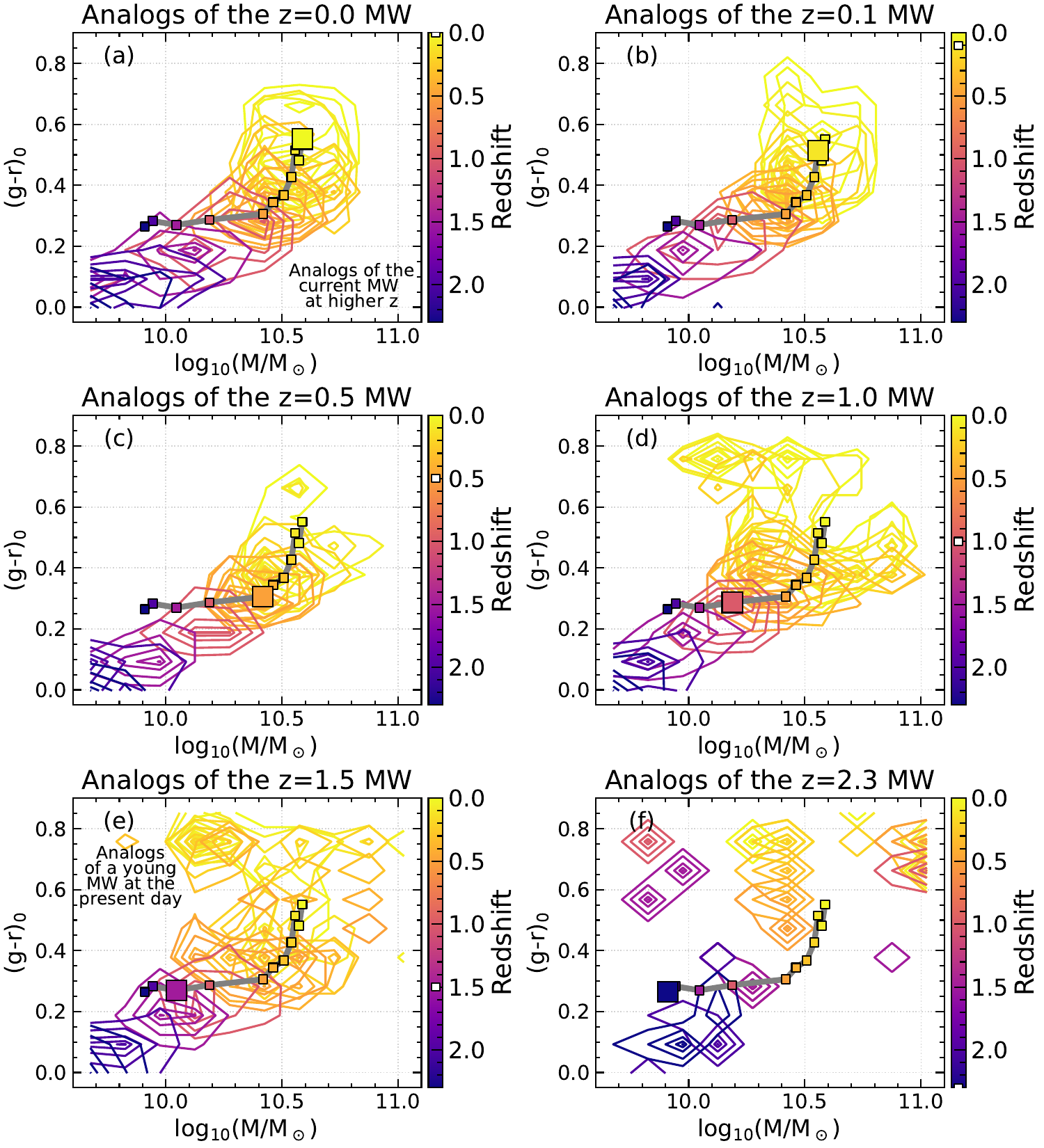}
    \caption{
    The evolving $M_*-(g-r)_0$ distributions of TNG50 ``Milky Way analogs'', selected as analogs at different points in the MW's history (\S\ref{sec:tng_description}). In each panel, the large square indicates our inferred MW mass and color at the given redshift (\S\ref{sec:time_machine_results}), and the contours describe the mass--color distributions of analogs defined at that redshift, as they appeared at earlier and later times as appropriate. All colors are scaled to the (identical) colorbars alongside each panel. See \S\ref{sec:tng_analog_paths} for details.
    }
    \label{fig:tng_compare}
\end{figure*}

The summary results of this analog analysis are shown in Figure~\ref{fig:tng_compare}. The small colored squares are identical in all panels and correspond to the MW's integrated values at each redshift indicated by the colorbar. Each panel highlights the selection of mass--color analogs of the MW {\it as it was at the given redshift}, as indicated by the panel's larger colored square (\S\ref{sec:tng_description}). The colored contours show, for the MW analogs selected at that redshift, how their mass--color distribution evolves at earlier and later times. 

\begin{figure*}[!ht]
    \centering
    \includegraphics[width=\textwidth]{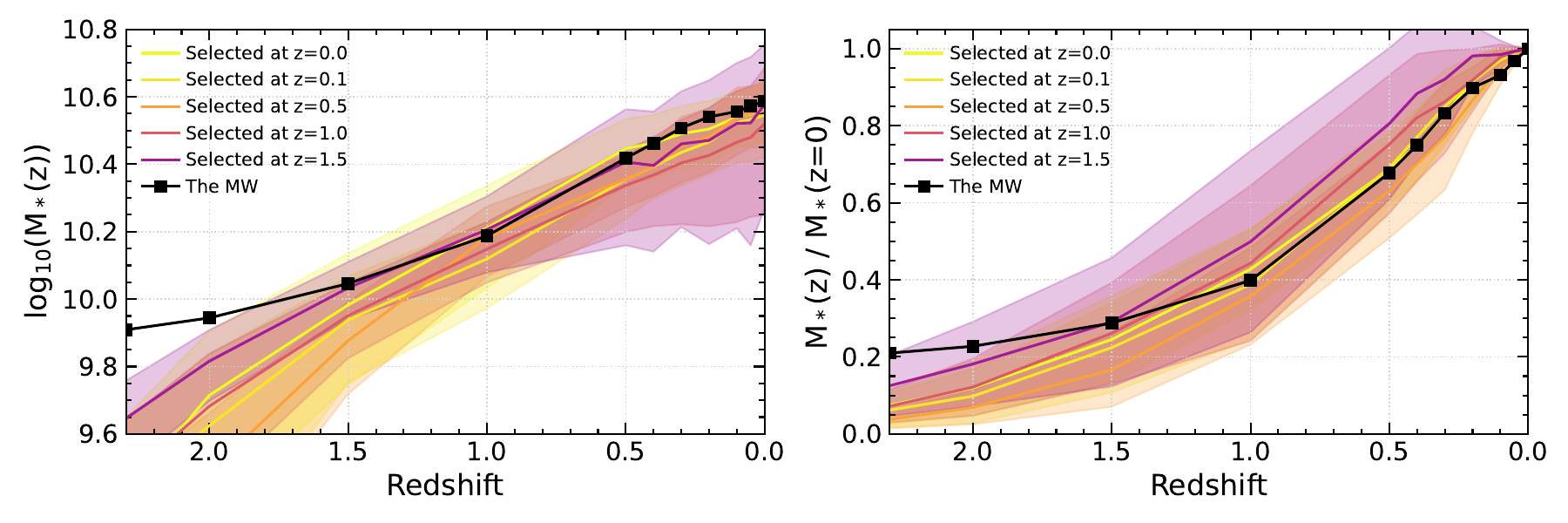}
    \caption{
    The mass assembly history of our inferred MW (black points) and its TNG50 analogs selected at different redshifts (\S\ref{sec:tng_description}). On the {\it left} is shown the buildup in absolute mass, and on the {\it right} is the cumulative assembly. See \S\ref{sec:tng_analog_paths} for details.
    }
    \label{fig:tng_mass_buildup}
\end{figure*}

For example, in Figure~\ref{fig:tng_compare}a, the upper left panel (``$z=0$''), the large yellow square indicates the MW at $z=0$, and the yellow contours indicate the mass--color distribution of subhalos whose $z=0$ properties match the $z=0$ MW. The other contours in that panel show the mass--color distribution of those same subhalos at earlier times, where the contour colors are described by the colorbar. Thus, simulated galaxies that are akin to the MW {\it today} were, for example, of similar mass but a few tenths of a dex bluer in $(g-r)_0$ back to $z \gtrsim 1$. In contrast, Figure~\ref{fig:tng_compare}e (``$z=1.5$'') shows in purple the integrated MW (square) and contemporary analog subhalos (contours), and yellow contours now indicate where those earlier analogs lie at $z=0$.

The MW and TNG50 galaxies' mass assembly histories are compared explicitly in Figure~\ref{fig:tng_mass_buildup}. The black line shows the MW's track, and each colored line shows the median subhalo stellar mass (and absolute median deviation of the masses, in the shaded region) of the analog systems selected at different redshifts. On the left is the absolute mass buildup (i.e., in $\log_{10}\left({M_*/M_\odot}\right)$), and on the right is the cumulative assembly, scaled by each system's stellar mass at $z=0$.

Between Figures~\ref{fig:tng_compare} and \ref{fig:tng_mass_buildup}, some general patterns can be identified:
\begin{itemize} \itemsep -2pt
    \item The slope of the MW's $dM_*/dz$ track is shallower than nearly all of its analogs.
    \item The {\it shape} of the MW's $M_*-(g-r)_0$ track is very similar to that of its $z=0$ analogs, but shifted to higher masses and redder colors at any given point in time (or, alternatively, reaching a given mass and redder color earlier in time than its $z=0$ analogs).
    It is possible that a small fraction of this color offset could be due to the lack of $\tau<10$~Myr MAAPs considered in our MW colors, but there is never enough mass that young to explain a majority of the offset.
    \item The spread, or diversity,  of analog masses and colors increases as we look at analogs defined at earlier and earlier times. But the reverse is not true --- analogs defined at $z=0$ are more tightly clumped at earlier times.
    \item Analogs of the early MW, especially above $z \gtrsim 1$, have a significant population of low-mass, very red systems today. 
\end{itemize}

The top-level picture that emerges from this novel comparison is that of a MW that assembled a larger fraction of its stellar mass prior to $z \sim 1-1.5$ ($7-9$~Gyr ago) than other MW-mass galaxies today, and has been producing or accreting fewer stars since then. That is, the vast majority of the TNG analogs of the present-day MW had assembled a much lower fraction of their stellar mass than the contemporary MW at redshifts earlier than $z \sim 1$; by definition, they then accumulated a larger fraction of their stellar mass more recently. 

This simulation-based narrative is entirely consistent with the early MW's evolution summarized in \S\ref{sec:intro} --- a stellar mass assembly more heavily weighted to higher redshifts, followed by a more quiescent period, than its present-day counterparts \citep[defined in a variety of ways;][]{Hammer_2007_MWtooquiet,Mackereth_2018_simalphaelements,Belokurov_2022_aurora,Rix_2022_pooroldheart,Chandra_2024_3phaseMW}. 
\textcolor{black}{One analogy that comes to mind is that of a wildly popular child movie star who avoided a self-destructive flame-out period and now lives as an ordinary, inconspicuous adult.}
The study by \citet{Martig_2021_earlydisc} of NGC~5746, which revealed a similarly early stellar growth history for a similarly massive disk galaxy, with a similarly subdominant classical bulge \citep[e.g.,][]{Shen_10_purediskbulge,BlandHawthorn_2016_MWreview}, presents an intriguing real-world analogy.

Additional analyses of simulations point to the potential impact of environment on this timeline. For example, \citet{Santistevan_2020_protoMWfire} and \citet{Horta_2024_protoMWfire} used the FIRE-2 simulations to argue that the progenitors of present-day MW-like galaxies emerged earlier when in Local Group-like environments, compared to isolated environments. We do not go into that level of detail with our ``early MW'' analogs here, which would be an interesting follow-up study.

Relatedly, though, we return to one of the original motivations for this study: not only how unique is the MW today, but also how unique has its evolutionary history been? The final two bullet points above relate to this question. Analogs of the present-day MW have followed a roughly consistent path through time --- at least, consistent with each other, if slightly shifted from the MW's path. In contrast, analogs of the much younger MW took a wide variety of paths as the Universe aged. 
In particular, a large fraction of the $z>1$ analogs have already, today, evolved through the green valley to lower masses and redder colors than the MW. For example, of the analog halos of the $z=1.5$~MW (panel~e in Figure~\ref{fig:tng_compare}), 41\% are smaller and redder than the MW at $z=0$, and only 28\% would still (at $z=0$) be considered ``analogs'' under the definition we have used here.
Essentially, these systems underwent a much more dramatic cessation of star formation than the MW, and moved nearly vertically through the green valley. A smaller, but nonzero, number of systems are both redder and more massive. Of course, an enormous amount of work has been done on quenching mechanisms and galaxy color evolution \citep[e.g.,][]{Faber_2007_redgals,Martig_2009_etgs,ForsterSchreiber_2020_cosmicnoonSF}; what is added in this study is the inference that at least some of these systems once looked as the MW did, which provides a very interesting additional constraint on the influence of different galaxy properties on these types of evolutionary processes.

\section{Summary and Conclusions}
\label{sec:conclusions}

We approach the long-standing challenge of characterizing the Milky Way Galaxy as ``a lower-case-g galaxy'' using a novel method based on widespread mapping of its mono-age-and-abundance populations (MAAPs). The original stellar density MAAPs, based on APOGEE data and corrected for the survey selection function, are constructed by \imig and span nearly 1.5~dex in metallicity and 13~Gyr in age (encompassing $\gtrsim$95\% of the stars in the MW; \S\ref{sec:maaps}). We describe modifications to the MAAPs to include younger populations ($\tau < 1$~Gyr); to extrapolate, interpolate, and smooth the inferred MAAP masses to reduce the impact of unconverged fits and other numerical artifacts; and to impose a physically motivated finer age-binning scheme (\S\ref{sec:maap_mods}). The mass, age, and metallicity of each MAAP is used to calculate its intrinsic luminosity in several bandpasses (\S\ref{sec:maap_mags}; these are then combined with a realistic dust model (\S\ref{sec:reddening}) to produce both intrinsic and reddened integrated colors of the MW (\S\ref{sec:combine_mags}).

We then explore numerous applications of this approach, including the construction of mass- and light-weighted metallicity distribution functions (MDFs; \S\ref{sec:mdfs_and_gradients}) and star formation histories (SFHs; \S\ref{sec:sfhs}). We find generally good agreement with other measures of these quantities where comparable values are available, but using the MAAPs, these distributions can be computed for arbitrary regions of the MW.

Taking advantage of the mono-age aspect of the MAAPs, we then ``run the clock backwards'' to calculate the MW's stellar mass, integrated colors, mean metallicities, and other properties at earlier times in its history (\S\ref{sec:time_machine_method}). We estimate that the MW has been a green valley galaxy for only a Gyr or two, and we show how its mass-weighted integrated stellar MDF has evolved over time through the multiple star formation episodes at different locations in the Galaxy (\S\ref{sec:time_machine_results}). We compare the time-resolved track of the MW in the mass--metallicity plane to several statistical stellar mass--metallicity relations derived for extragalactic systems, and discuss how future analyses along these lines may provide insight on the impact of different quenching mechanisms in the MW (\S\ref{sec:time_machine_results}). 

We briefly investiage the potential impact of including the Gaia-Enceladus progenitor and the Sgr dwarf galaxy on our findings, and conclude they do not have an effect (\S\ref{sec:mergers}).

Finally, we turn to the TNG50 simulations to select mass--color analogs of the MW not only at the present day, but also at higher redshifts (\S\ref{sec:tng_description}). We find compelling evidence for a higher fraction of the MW's stellar mass being in place at earlier times than for many of its present-day analogs, in agreement with other studies (\S\ref{sec:tng_analog_paths}). This time offset aside, simulated galaxies that are akin to the MW today followed a similar {\it path} through this mass--color space. We also find that simulated analogs of the early MW --- again, here selected by {\it observationally-inferred} masses and colors of the MW at early times --- take a variety of paths that often lead to quite different un-MW-like outcomes at $z=0$. In short, the MW is a product of its unique history; its fate was not written in its stars at early times.

\section{Acknowledgments}
This work was supported by the National Science Foundation under Grant No. AST-2009993.

The authors are grateful to Robert Butler, Leo Girardi, Stephanie Monty, Dylan Nelson, Annalisa Pillepich, Sanjib Sharma, Juan Soler, Zixian Wang (Purmortal), and Andrew Wetzel for useful discussions, suggestions, and/or assistance with ancillary data. They would also like to thank the anonymous referee for comments and suggestions that improved the clarity of this manuscript.

Funding for the Sloan Digital Sky Survey IV has been provided by the Alfred P. Sloan Foundation, the U.S. Department of Energy Office of Science, and the Participating Institutions. 

SDSS-IV acknowledges support and resources from the Center for High Performance Computing  at the University of Utah. The SDSS website is \url{www.sdss4.org}.

SDSS-IV is managed by the Astrophysical Research Consortium for the Participating Institutions of the SDSS Collaboration including the Brazilian Participation Group, the Carnegie Institution for Science, Carnegie Mellon University, Center for Astrophysics | Harvard \& Smithsonian, the Chilean Participation Group, the French Participation Group, Instituto de Astrof\'isica de Canarias, The Johns Hopkins University, Kavli Institute for the Physics and Mathematics of the Universe (IPMU) / University of Tokyo, the Korean Participation Group, Lawrence Berkeley National Laboratory, Leibniz Institut f\"ur Astrophysik Potsdam (AIP),  Max-Planck-Institut f\"ur Astronomie (MPIA Heidelberg), Max-Planck-Institut f\"ur Astrophysik (MPA Garching), Max-Planck-Institut f\"ur Extraterrestrische Physik (MPE), National Astronomical Observatories of China, New Mexico State University, New York University, University of Notre Dame, Observat\'ario Nacional / MCTI, The Ohio State University, Pennsylvania State University, Shanghai Astronomical Observatory, United Kingdom Participation Group, Universidad Nacional Aut\'onoma de M\'exico, University of Arizona, University of Colorado Boulder, University of Oxford, University of Portsmouth, University of Utah, University of Virginia, University of Washington, University of Wisconsin, Vanderbilt University, and Yale University.

The IllustrisTNG simulations were undertaken with compute time awarded by the Gauss Centre for Supercomputing (GCS) under GCS Large-Scale Projects GCS-ILLU and GCS-DWAR on the GCS share of the supercomputer Hazel Hen at the High Performance Computing Center Stuttgart (HLRS), as well as on the machines of the Max Planck Computing and Data Facility (MPCDF) in Garching, Germany.


%

\vspace{5mm}
\facilities{Du Pont (APOGEE), Sloan (APOGEE)}


\software{astropy \citep{astropy_2013,astropy_2018,astropy_2022}, scipy \citep{scipy_2020}, kcorrect \citep{Blanton_2007_kcorrect}, matplotlib \citep{Hunter_2007_mpl}}



\bibliography{ms}{}
\bibliographystyle{aasjournal}

\end{document}